\documentclass[fleqn,10pt]{revtex4}

\usepackage{amsmath}
\usepackage{amssymb}
\usepackage{graphicx,psfrag}
\usepackage{afterpage}
\usepackage{mathrsfs}

\DeclareMathOperator{\tr}{Tr}

\def\slash#1{\setbox0=\hbox{$#1$}               % set a box for #1
        \dimen0=\wd0                            % and get its size
        \setbox1=\hbox{/} \dimen1=\wd1          % get size of /
        \ifdim\dimen0>\dimen1                   % #1 is bigger
        \rlap{\hbox to \dimen0{\hfil/\hfil}}    % so center / in box
        #1                                      % and print #1
        \else                                   % / is bigger
        \rlap{\hbox to \dimen1{\hfil$#1$\hfil}} % so center #1
        /                                       % and print /
        \fi}                                    %

\begin{document}

\title{Non-universality of transverse momentum dependent parton 
distribution functions}

\author{C.J. Bomhof}
\email{cbomhof@nat.vu.nl}
\affiliation{
Department of Physics and Astronomy, Vrije Universiteit Amsterdam,\\
NL-1081 HV Amsterdam, the Netherlands}

\author{P.J. Mulders}
\email{mulders@few.vu.nl}
\affiliation{
Department of Physics and Astronomy, Vrije Universiteit Amsterdam,\\
NL-1081 HV Amsterdam, the Netherlands}

\begin{abstract}
In the field theoretical description of hadronic scattering processes,
single transverse-spin asymmetries arise due to gluon initial and final state interactions.
These interactions lead to process dependent Wilson lines in the operator definitions of transverse momentum dependent parton distribution functions.
In particular for hadron-hadron scattering processes with hadronic final states this has important ramifications for possible factorization formulas in terms of (non)universal TMD parton distribution functions.
In this paper we will systematically separate the universality-breaking parts of the TMD parton correlators from the universal $T$-even and $T$-odd parts.
This might play an important role in future factorization studies for these processes.
We also show that such factorization theorems will (amongst others) involve the gluonic pole cross sections,
which have previously been shown to describe the hard partonic scattering in weighted spin asymmetries. 
\end{abstract}
\date{\today}
%\pacs{12.38.-t; 13.85.Ni; 13.88.+e}

\maketitle

\section{Introduction}

Many theoretical as well as experimental studies in recent years have been aimed at better understanding the processes that cause spin asymmetries in hadronic scattering.
A mechanism to generate single-spin asymmetries (SSA) through soft gluon 
interactions between the target remnants and the initial and final state 
partons was first proposed in the context of collinear 
factorization~\cite{Efremov:1981sh,Efremov:1984ip,Qiu:1991pp,Qiu:1991wg,Qiu:1998ia,Kanazawa:2000hz,Eguchi:2006mc,Koike:2006qv}.
This collinear factorization formalism involves, 
apart from the usual twist-two quark correlators, 
also twist-three collinear quark-gluon matrix elements.
Since they contain the field operator of a zero-momentum gluon, 
they are referred to as \emph{gluonic pole matrix elements}.
An important example is the Qiu-Sterman matrix element $T_F(x{,}x)$~\cite{Efremov:1981sh,Efremov:1984ip,Qiu:1991pp,Qiu:1991wg,Qiu:1998ia}.

Several other mechanisms to generate SSA's through the effects of the 
intrinsic transverse momenta of the partons have also been proposed.
For instance, in the Sivers effect the asymmetry arises in the initial 
state due to a correlation between the intrinsic transverse motion of 
an unpolarized quark and the transverse spin of its parent 
hadron~\cite{Sivers:1989cc,Sivers:1990fh}.
The effect can be described by a transverse momentum dependent (TMD) distribution function $f_{1T}^\perp(x{,}p_T^2)$~\cite{Boer:1997nt}.
Such a function can exist by the grace of soft gluon interactions between the target remnants and the active partons~\cite{Brodsky:2002cx,Brodsky:2002rv}.
These interactions give rise to process dependent Wilson lines,
also called gauge links, 
in the definitions of TMD parton distribution and fragmentation functions.
The Wilson lines secure the gauge invariance of these definitions. 
At the same time they prevent the use of time-reversal to argue that the Sivers effect should vanish.
Instead, time-reversal can be used to derive non-trivial `universality' relations between the Sivers functions in different processes.
For instance, it was shown that the Sivers function in SIDIS,
which contains a future pointing Wilson line,
has opposite sign~\cite{Brodsky:2002cx,Brodsky:2002rv,Collins:2002kn} as the TMD function in Drell-Yan scattering,
which involves a past pointing Wilson line.
Moreover, the Wilson lines are also crucial ingredients in the derivation~\cite{Boer:2003cm} of the relation between the Sivers function and the Qiu-Sterman matrix element, 
$2Mf_{1T}^{\perp(1)}(x)\,{=}\,{-}gT_F(x{,}x)$, 
demonstrating that the first transverse moment of the Sivers function 
is a gluonic pole matrix element.

The process dependence of the Wilson lines in TMD parton correlators makes the study of the (non)universality of these functions particularly important.
In the basic electroweak processes, SIDIS, Drell-Yan scattering and 
$e^+e^-$-annihilation, 
the hard partonic parts of the process are just simple 
electroweak vertices (at tree-level).
Depending on the particular process only initial or final-state 
gluon interactions contribute and, correspondingly,
only future and past pointing Wilson lines occur.
However, when going to hadronic processes that involve hard parts 
with more colored external legs,
such as in hadronic dijet or photon-jet production,
there can be both initial and final state gluon interactions.
As a result, the Wilson lines resulting from a resummation of all exchanged collinear gluons will also be more complicated than just 
the simple future and past pointing Wilson 
lines~\cite{Bomhof:2004aw,Bacchetta:2005rm,Bomhof:2006dp}.
In particular, for each of the Feynman diagrams that contribute to the hard 
partonic part of the hadronic scattering process there is, in principle, 
a different gauge link structure.

For the TMD distribution functions this at first sight seems to complicate things considerably. 
However, for the collinear distribution functions remarkable simplifications 
occur. Upon integration over intrinsic transverse momenta all the effects of 
the complicated gauge link structures in the TMD correlators disappear,
while for the transverse moment they contribute a gluonic pole 
matrix element with multiplicative prefactors, 
referred to as \emph{gluonic pole strengths}. 
These are color-fractions that, in principle, 
differ for each Feynman diagram that contributes to the partonic subprocess. 
Therefore, for a given subprocess one can multiply the color factors with the contribution of each partonic diagram and collect them in modified 
(but manifestly gauge invariant) 
hard cross sections~\cite{Bacchetta:2005rm,Bomhof:2006ra}. 
These modified hard functions, called \emph{gluonic pole cross sections}, 
appear whenever gluonic pole matrix elements 
(such as the first moments of the Sivers and Boer-Mulders functions) contribute.
This is typically the case in weighted azimuthal spin asymmetries.
%Several other recent theoretical studies of single-spin asymmetries 
%point in a similar 
%direction~\cite{Koike:2007rq,Ratcliffe:2007ye,Qiu:2007ar,Qiu:2007ey}, 
%though more research is required to convey the exact connection between 
%the different formalisms.

The effects of the gluon initial and final-state interactions for the 
fully TMD treatment of these processes is less clear-cut.
In Refs~\cite{Qiu:2007ar,Qiu:2007ey} a TMD factorization formula based on one-gluon radiation was proposed for the quark-Sivers contribution to the SSA in dijet production in proton-proton scattering. 
This result involves the gluonic pole cross sections found in Refs~\cite{Bacchetta:2005rm,Bomhof:2006ra} as hard parts, 
folded with the TMD distribution functions as measured in SIDIS
(\emph{i.e.}\ with a future pointing Wilson line in their definitions).
On the other hand, in Refs~\cite{Bomhof:2004aw,Bacchetta:2005rm,Bomhof:2006dp} 
it was observed that complicated Wilson line structures occur in the TMD 
distribution (and fragmentation) functions in such processes.
Those results, in concurrence with a model calculation,
led the authors of Ref.~\cite{Collins:2007nk} to conclude that a TMD 
factorization formula for spin asymmetries in processes such as dijet 
production in proton-proton scattering cannot be written down 
with universal distribution functions.
It is also asserted that a proof of TMD factorization for such processes 
will be essentially different from the existing proofs for SIDIS and 
Drell-Yan scattering and that it will probably involve `effective' TMD parton distribution functions~\cite{Ratcliffe:2007ye}.
%Very recently it has been argued in %Refs~\cite{Vogelsang:2007jk,Collins:2007jp} that the work %of~\cite{Bomhof:2004aw,Bacchetta:2005rm,Bomhof:2006dp,Bomhof:2006ra},
%\cite{Qiu:2007ar,Qiu:2007ey} and~\cite{Collins:2007nk} are all consistent to at %least the two-gluon exchange contributions.
Recent extensions~\cite{Vogelsang:2007jk,Collins:2007jp} of the work in~\cite{Qiu:2007ar,Qiu:2007ey,Collins:2007nk} also include the contributions of two collinear gluons
(as was previously discussed for Drell-Yan~\cite{Boer:1999si}).
These indicate that the Feynman graph calculations in 
Refs~\cite{Bomhof:2004aw,Bacchetta:2005rm,Bomhof:2006dp,Bomhof:2006ra},
\cite{Qiu:2007ar,Qiu:2007ey} and~\cite{Collins:2007nk} are ``mutually consistent" up to two-gluon contributions~\cite{Vogelsang:2007jk}.

By using the gluonic pole strengths we will in this paper  
systematically separate the universality-breaking parts 
of the TMD parton correlators from the universal $T$-even 
and $T$-odd matrix elements.
It is a non-trivial observation that this is possible and we believe 
that it constitutes another important ingredient in trying to relate the results of Refs~\cite{Bomhof:2004aw,Bacchetta:2005rm,Bomhof:2006dp,Bomhof:2006ra} 
and Refs~\cite{Qiu:2007ar,Qiu:2007ey}.
We demonstrate that the gluonic pole cross sections are also encountered in unintegrated, unweighted processes.
In particular, we will argue that the gluonic pole cross sections can also emerge in unweighted spin-averaged processes and that ordinary partonic cross sections can also arise in unweighted single-spin asymmetries,
though they appear in such a way that they will vanish for the integrated and weighted processes~\cite{Bacchetta:2005rm,Bomhof:2006ra,Bomhof:2007su,Bacchetta:2007sz}, respectively.
We will start by recapitulating the collinear case in 
section~\ref{CollSection} and the appearance of universal collinear
correlators in hadronic cross sections  in section~\ref{Boterham}.  
The study of the non-universality of the 
TMD parton correlators will be presented in sections~\ref{TMDSection} and~\ref{Parameterizations}, 
followed by a discussion on how the non-universal TMD correlators 
affect hadronic cross sections (section~\ref{Beleg}).
After summarizing in section~\ref{Conclusion} we list all universality-breaking matrix elements that are encountered at tree-level in $2{\rightarrow}2$ hadronic scattering processes (appendix~\ref{CORRS}).

\section{Collinear Correlators\label{CollSection}}

For a twist analysis of hadronic variables in high-energy physics it is 
useful to make a Sudakov decomposition 
$p^\mu\,{=}\,xP^\mu{+}\sigma n^\mu{+}p_T^\mu$
of the momentum $p^\mu$ of each active parton.
The Sudakov vector $n$ is an arbitrary light-like four-vector $n^2\,{\equiv}\,0$ that has non-zero overlap $P{\cdot}n$ with the hadron's momentum $P^\mu$.
We will choose the Sudakov vector such that this overlap is positive and of the order of the hard scale.
Up to subleading twist its coefficient 
$\sigma\,{=}\,(p{\cdot}P{-}xM^2)/(P{\cdot}n)$ is always integrated over.
%and will only appear explicitly in terms that are suppressed by two powers of %the hard scale as compared to the collinear term.}
The vector $p_T$ is called the intrinsic transverse momentum of the parton. 
It is orthogonal to both $P$ and $n$,
\emph{i.e.}~$p_T{\cdot}P\,{=}\,p_T{\cdot}n\,{=}\,0$,
and will appear suppressed by one power of the hard scale with respect to the 
collinear term.
Vectors in the transverse plane can be obtained by using the transverse 
projectors 
$g_T^{\mu\nu}\,{\equiv}\,g^{\mu\nu}{-}P^{\{\mu}n^{\nu\}}/(P{\cdot}n)$ 
and 
$\epsilon_T^{\mu\nu}\,
{\equiv}\,\epsilon^{\mu\nu\rho\sigma}P_\rho n_\sigma/(P{\cdot}n)$.
Note that each observed hadron can have a different transverse plane.

\begin{figure}
\centering
\begin{minipage}{3.8cm}
\centering
\includegraphics[width=\textwidth]{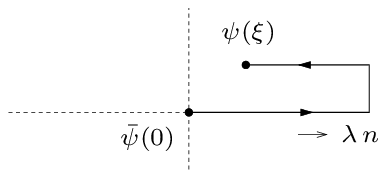}\\%[1mm]
$\scriptstyle{\Phi^{[+]}{\propto}
\langle\bar\psi(0)\mathcal U^{[+]}\psi(\xi)\rangle}$\\[2mm]
(a)
\end{minipage}
\hspace{0.5cm}
%%%%
%%%%
\begin{minipage}{3.8cm}
\centering
\includegraphics[width=\textwidth]{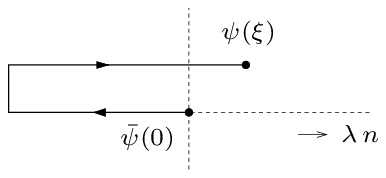}\\%[1mm]
$\scriptstyle{\Phi^{[-]}{\propto}
\langle\bar\psi(0)\mathcal U^{[-]}\psi(\xi)\rangle}$\\[2mm]
(b)
\end{minipage}\\[4mm]
%%%%
%%%%
\begin{minipage}{3.8cm}
\centering
\includegraphics[width=\textwidth]{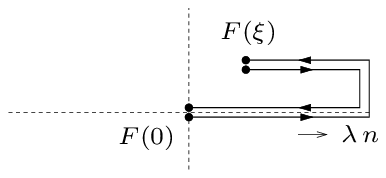}\\%[1mm]
$\scriptstyle{\Gamma^{[+,+^\dagger]}{\propto}
\tr\langle F(0)\mathcal U^{[+]}F(\xi)\mathcal U^{[+]\dagger}\rangle}$\\[2mm]
(c)
\end{minipage}
\hspace{0.5cm}
%%%%
%%%%
\begin{minipage}{3.8cm}
\centering
\includegraphics[width=\textwidth]{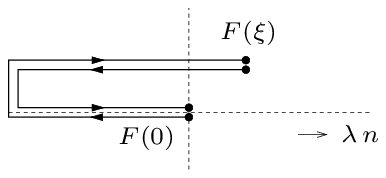}\\%[1mm]
$\scriptstyle{\Gamma^{[-,-^\dagger]}{\propto}
\tr\langle F(0)\mathcal U^{[-]}F(\xi)\mathcal U^{[-]\dagger}\rangle}$\\[2mm]
(d)
\end{minipage}
\hspace{0.5cm}
%%%%
%%%%
\begin{minipage}{3.8cm}
\centering
\includegraphics[width=\textwidth]{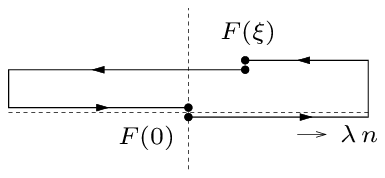}\\%[1mm]
$\scriptstyle{\Gamma^{[+,-^\dagger]}{\propto}
\tr\langle F(0)\mathcal U^{[+]}F(\xi)\mathcal U^{[-]\dagger}\rangle}$\\[2mm]
(e)
\end{minipage}
\hspace{0.5cm}
%%%%
%%%%
\begin{minipage}{3.8cm}
\centering
\includegraphics[width=\textwidth]{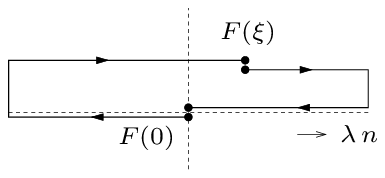}\\%[1mm]
$\scriptstyle{\Gamma^{[-,+^\dagger]}{\propto}
\tr\langle F(0)\mathcal U^{[-]}F(\xi)\mathcal U^{[+]\dagger}\rangle}$\\[2mm]
(f)
\end{minipage}
\parbox{0.95\textwidth}{\caption{
Simplest structures (without loops)
for gauge links and operators in quark correlators (a)-(b) 
and gluon correlators (c)-(f). 
\label{simplelinks}}}
\end{figure}

We consider collinear quark distribution functions as being obtained from transverse momentum dependent (TMD) quark distribution functions.
Those are projections of the TMD quark correlator defined on the light-front 
(LF: $\xi{\cdot}n\,{\equiv}\,0$)
\begin{equation}\label{TMDcorrelator}
\Phi_{ij}^{[\mathcal U]}(x{,}p_T{;}P{,}S)
={\int}\frac{d(\xi{\cdot}P)d^2\xi_T}{(2\pi)^3}\ e^{ip\cdot\xi}\,
\langle P{,}S|\,\overline\psi_j(0)\,\mathcal U_{[0;\xi]}\,
\psi_i(\xi)\,|P{,}S\rangle\big\rfloor_{\text{LF}}\ .
\end{equation}
The \emph{Wilson line} or \emph{gauge link}
$\mathcal U_{[\eta;\xi]}\,
{=}\,\mathcal P{\exp}\big[{-}ig{\int_C}\,ds{\cdot}A^a(s)\,t^a\,\big]$
is a path-ordered exponential along the integration path $C$ with 
endpoints at $\eta$ and $\xi$.
Its presence in the hadronic matrix element is required by gauge-invariance.
In the TMD correlator~\eqref{TMDcorrelator} the integration path $C$ in 
the gauge link is process-dependent.
In the diagrammatic approach the Wilson lines arise by resumming all 
gluon interactions between the soft and the hard partonic parts of the 
hadronic process~\cite{Efremov:1978xm,Boer:1999si,Belitsky:2002sm,Boer:2003cm}.
Consequently, the integration path $C$ is fixed by the (color-flow structure 
of) the hard partonic scattering~\cite{Bomhof:2006dp}.
Basic examples are semi-inclusive deep-inelastic scattering (SIDIS) where 
the resummation of all final-state interactions leads to the future 
pointing Wilson line $\mathcal U^{[+]}$,
and Drell-Yan scattering where the initial-state interactions lead to 
the past  pointing Wilson line $\mathcal U^{[-]}$,
see Figures~\ref{simplelinks}a and~b.
All Wilson lines in this text are in the three-dimensional fundamental 
representation of the color matrices.
%\begin{align}
%\mathcal U^{[\pm]}
%&=U_{[(0^-,\boldsymbol0_T);(\pm\infty^-,\boldsymbol0_T)]}^n
%U_{[(\pm\infty^-,\boldsymbol0_T);(\pm\infty^-,\boldsymbol\infty_T)]}^T
%U_{[(\pm\infty^-,\boldsymbol\infty_T);(\pm\infty^-,\boldsymbol\xi_T)]}^T
%U_{[(\pm\infty^-,\boldsymbol\xi_T);(\xi^-,\boldsymbol\xi_T)]}^n\ ,\label{GL}
%\end{align}
%where the $U_{[\eta{;}\xi]}$ are gauge links along straight lines:
%\begin{equation}
%U_{[\eta;\xi]}^n
%=\mathcal P\exp\Big[-ig\int_{n\cdot\eta}^{n\cdot\xi}ds\ 
%n{\cdot}A^a(s)\,t^a\,\Big]\ ,
%%%%%
%\qquad U_{[\eta;\xi]}^T
%=\mathcal P\exp\Big[-ig\int_{\eta_T}^{\xi_T}ds_T\cdot
%A_T^a(s)\,t_{\phantom{T}}^a\,\Big]\ .
%\end{equation}

Going beyond the simplest electroweak processes such as SIDIS, 
Drell-Yan scattering and $e^+e^-$-annihilation, the competing effects 
of the gluonic initial and final-state interactions lead 
to gauge link structures that can be quite more complicated than the 
future or past pointing Wilson lines~\cite{Bomhof:2004aw,Bacchetta:2005rm,Bomhof:2006dp}.
The situation becomes particularly notorious when considering processes 
which have several Feynman diagrams that contribute to the partonic scattering. 
In that case each cut Feynman diagram $D$ can, in principle,
lead to a different gauge-invariant correlator
$\Phi^{[\mathcal U]}(x{,}p_T)\,{=}\,\Phi^{[\mathcal U(D)]}(x{,}p_T)\,
{\equiv}\,\Phi^{[D]}(x{,}p_T)$~\cite{Bomhof:2004aw,Bacchetta:2005rm,Bomhof:2006dp}.
This observation leads to a broad spectrum of different TMD parton 
correlators that appear in hadronic scattering processes.

For collinear correlators the situation is simpler.
For instance, in the $p_T$-integrated correlator defined on the 
lightcone (LC: $\xi{\cdot}n\,{=}\,\xi_T\,{\equiv}\,0$)
\begin{subequations}
\begin{align}
\Phi_{ij}(x)
&={\int}d^2p_T\ \;\Phi_{ij}^{[\mathcal U]}(x{,}p_T)
={\int}\frac{d(\xi{\cdot}P)}{2\pi}\ e^{ip{\cdot}\xi}
\langle P{,}S|\,\overline\psi_j(0)\,U_{[0;\xi]}^n\,
\psi_i(\xi)\,|P{,}S\rangle\big\rfloor_{\text{LC}}\ ,
\end{align}
\end{subequations}
all process-dependence of the gauge link disappears,
leaving just a straight Wilson line $U_{[0;\xi]}^n$ in the 
lightcone $n$-direction,
where $n$ is the lightlike vector in the Sudakov decomposition of the 
quark momentum $p$ (we will use the non-calligraphed letter $U$ to indicate 
straight line segments).
Another situation is encountered in the transverse momentum 
weighted correlators (the transverse moments). 
In the transverse moments a (sub)process-dependence remains as a direct consequence of the presence of the gauge links in the TMD correlators. 
Nevertheless, a simple decomposition can still be made
(omitting the Dirac indices)~\cite{Boer:2003cm,Bacchetta:2005rm,Bomhof:2006ra}:
\begin{equation}\label{TransverseMomentGeneral}
\Phi_\partial^{[\mathcal U]\,\alpha}(x)
\equiv{\int}d^2p_T^{\phantom{\alpha}}\ \;p_T^\alpha\ \;
\Phi^{[\mathcal U]}(x{,}p_T)
=\widetilde\Phi{}_\partial^\alpha(x)
+C_G^{[\mathcal U]}\,\pi\Phi_G^\alpha(x{,}x)\ ,
\end{equation}
with collinear correlators 
%($F^{n\alpha}\,{\equiv}\,F^{\mu\alpha}n_\mu/(P{\cdot}n)$)
\begin{subequations}
\begin{gather}
\Phi_D^\alpha(x)
=\int\frac{d(\xi{\cdot}P)}{2\pi}\ e^{ix(\xi\cdot P)}\,
\langle P{,}S|\,\overline\psi(0)\,U_{[0;\xi]}^n\,iD^\alpha(\xi)\,
\psi(\xi)\,|P{,}S\rangle\,\big\rfloor_{\text{LC}}\ ,\\
%%%%
\Phi_G^\alpha(x,x{-}x')
=\frac{n_\mu}{P{\cdot}n}
\int\frac{d(\xi{\cdot}P)}{2\pi}\frac{d(\eta{\cdot}P)}{2\pi}\ 
e^{ix'(\eta\cdot P)}e^{i(x-x')(\xi\cdot P)}\,
\langle P{,}S|\,\overline\psi(0)\,U_{[0;\eta]}^n\,gF^{\mu\alpha}(\eta)\,
U_{[\eta;\xi]}^n\,\psi(\xi)\,|P{,}S\rangle\,\big\rfloor_{\text{LC}}\ ,
\end{gather}
\end{subequations}
and 
\begin{equation}
\widetilde\Phi{}_\partial^\alpha(x)
=\Phi_D^\alpha(x)
-\int dx'\ P\frac{i}{x'}\ \Phi_G^\alpha(x,x{-}x')\ .\\
\end{equation}
The only process dependence due to the Wilson lines in the 
TMD correlators resides in the multiplicative factors 
$C_G^{[\mathcal U]}\,{=}\,C_G^{[\mathcal U(D)]}\,{\equiv}\,C_G^{[D]}$.
They are color-fractions that are fixed by the color-flow structure of the hard 
partonic function of the scattering 
process~\cite{Bacchetta:2005rm,Bomhof:2006ra}.
We will refer to them as \emph{gluonic pole strengths}.
Important examples are the
transverse moments of the correlators $\Phi^{[+]}$ in SIDIS and 
$\Phi^{[-]}$ in Drell-Yan scattering, 
for which one has
$C_G^{[\pm]}\,
{\equiv}\,C_G^{[\mathcal U^{[\pm]}]}\,{=}\,{\pm}1$~\cite{Boer:2003cm}.

Transverse momentum dependent gluon distribution functions are projections of the TMD correlator
\begin{equation}\label{GluonTMD}
%\Phi_{(g)}^{[\mathcal U,\mathcal U^\prime]\,\mu\nu}(x{,}p_T)
\Gamma^{[\mathcal U,\mathcal U^\prime]\,\mu\nu}(x{,}p_T{;}P{,}S)
=\frac{n_\rho n_\sigma}{(p{\cdot}n)^2}
{\int}\frac{d(\xi{\cdot}P)\;d^2\xi_T}{(2\pi)^3}\ e^{ip\cdot\xi}
\tr\,\langle P{,}S|\,F^{\mu\rho}(0)\,
\mathcal U_{[0{;}\xi]}^{\phantom{\prime}}\,
F^{\nu\sigma}(\xi)\,\mathcal U_{[\xi{;}0]}^\prime\,
|P{,}S\rangle\,\big\rfloor_{\text{LF}}\ .
\end{equation}
Here $\tr$ indicates a trace over color-triplet indices.
Writing the field-operators in the color-triplet representation requires the inclusion of \emph{two} Wilson lines $\mathcal U_{[0{;}\xi]}$ and 
$\mathcal U_{[\xi{;}0]}^\prime$~\cite{Bomhof:2006dp}.
They again arise from the resummation of gluon initial and final-state 
interactions. In general this will lead to two unrelated Wilson 
lines $\mathcal U$ and $\mathcal U'$.
In the particular case that $\mathcal U'\,{=}\,\mathcal U^\dagger$,
the gluon correlator can also be written as the product
$\langle F_a\mathcal U_{ab}F_b\rangle$ of two gluon fields 
with the Wilson line $\mathcal U$ in the adjoint representation of $SU(N)$.
This is for instance the case for the gluon correlators in 
Figs~\ref{simplelinks}c and~d,
but not for the gluon correlators in Figs~\ref{simplelinks}e and~f.

In the $p_T$-integrated correlator on the lightcone the
process dependence of the TMD gluon correlator disappears,
\begin{equation}\label{Gluon}
\Gamma^{\mu\nu}(x)
={\int}d^2p_T\ \ \Gamma^{[\mathcal U{,}\mathcal U']\;\mu\nu}(x{,}p_T)
=\frac{n_\rho n_\sigma}{(p{\cdot}n)^2}
{\int}\frac{d(\xi{\cdot}P)}{2\pi}\ e^{ix(\xi\cdot P)}\,
\tr\,\langle P{,}S\vert\,F^{\mu\rho}(0)\,U_{[0;\xi]}^n\,
F^{\nu\sigma}(\xi)\,U_{[\xi;0]}^n\vert P{,}S\rangle\,\big\rfloor_{\text{LC}}\ .
\end{equation}
However, as for the quark correlator,
a subprocess-dependence due to the Wilson lines in the TMD gluon 
correlators remains in the transverse moments.
The analogue of the decomposition~\eqref{TransverseMomentGeneral} 
in the case of the gluon correlator is
(with $\Gamma^{[\mathcal U,\mathcal U^\prime]}(x{,}p_T)\,
{=}\,\Gamma^{[\mathcal U(D),\mathcal U^\prime(D)]}(x{,}p_T)\,
{\equiv}\,\Gamma^{[D]}(x{,}p_T)$ and omitting the gluon field indices 
$\mu$ and $\nu$)~\cite{Bomhof:2006ra}:
\begin{equation}\label{GluonDecomp}
\Gamma_{\partial}^{[D]\,\alpha}(x)
=\widetilde\Gamma_{\partial}^\alpha(x)
+C_G^{(f)\,[D]}\,\pi\Gamma_{G_f}^\alpha(x{,}x)
+C_G^{(d)\,[D]}\,\pi\Gamma_{G_d}^\alpha(x{,}x)\ .
\end{equation}
The matrix elements $\Gamma_{G_f}$ and $\Gamma_{G_d}$ are 
the two gluonic pole matrix elements that correspond to the two possible ways 
to construct color-singlets from three gluon 
fields~\cite{Ji:1992eu,Bomhof:2006ra}.
They involve the antisymmetric $f$ and symmetric $d$ structure
constants of $SU(3)$, respectively. 
The only process dependence coming from the Wilson lines in the TMD 
correlators is contained in the gluonic pole strengths
$C_G^{(f/d)\,[D]}\,
{\equiv}\ C_G^{(f/d)\,[\mathcal U(D),\mathcal U^\prime(D)]}$. 
The collinear correlators are
\begin{subequations}
\begin{gather}
\Gamma_D^{\mu\nu;\alpha}(x)
=\frac{n_\rho n_\sigma}{(p{\cdot}n)^2}
\int\frac{d(\xi{\cdot}P)}{2\pi}\ e^{ix(\xi{\cdot}P)}\,
\tr\,\langle P{,}S|\,F^{\mu\rho}(0)\,U^n_{[0;\xi]}\,
\big[iD^\alpha(\xi),F^{\nu\sigma}(\xi)\big]\,
U^n_{[\xi;0]}\,|P{,}S\rangle\,\big\rfloor_{\text{LC}}\ ,\\
%%%%
\Gamma_{G_f}^{\mu\nu;\alpha}(x,x{-}x')
=\frac{n_\rho n_\sigma}{(p{\cdot}n)^2}
\int\frac{d(\xi{\cdot}P)}{2\pi}\frac{d(\eta{\cdot}P)}{2\pi}\ 
e^{ix'(\eta{\cdot}P)}e^{i(x{-}x')(\xi{\cdot}P)}\nonumber\\
\mspace{260mu}\times\ 
\tr\,\langle P{,}S|\,F^{\mu\rho}(0)\,
\big[U_{[0,\eta]}^ngF^{n\alpha}(\eta)U_{[\eta,0]}^n,
U_{[0,\xi]}^nF^{\nu\sigma}(\xi)U_{[\xi,0]}^n\big]\,
|P{,}S\rangle\,\big\rfloor_{\text{LC}}\ ,
\label{sinaasappelsap}\displaybreak[0]\\
%%%%
\Gamma_{G_d}^{\mu\nu;\alpha}(x,x{-}x')
=\frac{n_\rho n_\sigma}{(p{\cdot}n)^2}
\int\frac{d(\xi{\cdot}P)}{2\pi}\frac{d(\eta{\cdot}P)}{2\pi}\ 
e^{ix'(\eta{\cdot}P)}e^{i(x{-}x')(\xi{\cdot}P)}\nonumber\\
\mspace{260mu}\times\ 
\tr\,\langle P{,}S|\,F^{\mu\rho}(0)\,
\big\{U_{[0,\eta]}^ngF^{n\alpha}(\eta)U_{[\eta,0]}^n,
U_{[0,\xi]}^nF^{\nu\sigma}(\xi)U_{[\xi,0]}^n\big\}\,
|P{,}S\rangle\,\big\rfloor_{\text{LC}}\ ,
\end{gather}
\end{subequations}
and 
\begin{equation}
\widetilde\Gamma{}_\partial^\alpha(x)
=\Gamma_D^\alpha(x)
-\int dx'\ P\frac{i}{x'}\ \Gamma_{G_f}^\alpha(x,x{-}x')\ .
\end{equation}
The collinear (anti)quark and gluon fragmentation correlators 
can be analyzed in the same way.
The matrix elements in~\eqref{TransverseMomentGeneral} 
and~\eqref{GluonDecomp} contain the collinear $T$-even and $T$-odd 
parton distribution functions, see \emph{e.g.}~\cite{Bomhof:2006ra}.

\section{Collinear functions in hadronic cross sections\label{Boterham}}

In the diagrammatic approach, the calculation of the hadronic cross 
sections starts off with the transverse momentum dependent 
parton correlators (TMD distribution and fragmentation functions), which
will appear in combination with squares of hard partonic amplitudes. 
In general the hard amplitude contains more terms, that is $H\,{=}\,\sum_iH_i$.
In the squared amplitude one therefore has terms like
$H_i^*H_j^{\phantom{*}}\,{\equiv}\,\hat\Sigma^{[D]}$,
where $D$ refers to the cut Feynman diagram that is the pictorial 
representation of the product of the amplitude $H_j$ and conjugate 
hard amplitude $H_i^*$. 
The hadronic cross section $d\sigma$ of a hadronic scattering process mediated by a two-to-two partonic subprocess
$a(p_1)b(p_2){\rightarrow}c(k_1)d(k_2)$ and where the outgoing hadrons and/or jets are produced with large perpendicular component with respect to the beam
will contain the following structure in the integrand:
\begin{equation}\label{shorthand}
\Sigma(p_1{,}p_2{,}k_1{,}k_2)
\equiv\sum_{a{,}{\cdots}{,}d}\sum_{D}
\Phi_a^{[D]}(x_1{,}p_{1T})\otimes\Phi_b^{[D]}(x_2{,}p_{2T})\otimes
\hat\Sigma^{[D]}(p_1{,}p_2{,}k_1{,}k_2)\otimes
\Delta_c^{[D]}(z_1{,}k_{1T})\otimes\Delta_d^{[D]}(z_2{,}k_{2T})\ ,
\end{equation}
where the parton momenta are approximately 
(compared to the hard scale) on-shell.
The convolutions `$\otimes$' represent the appropriate Dirac and color 
traces for the hard function $\hat\Sigma^{[D]}$.
To get to the hadronic cross section one has to multiply by the flux factor and integrate over the final-state phase-space and parton momenta including a delta function for momentum conservation on the partonic level.
Since our aim in this paper is to display some general features that a $k_T$-factorization formula (if it exists) will have due to the process-dependence of the Wilson lines that arise in the diagrammatic TMD gauge link approach,
we focus our discussion on the Wilson lines and neglect soft factors.
In the full $k_T$-factorization formula such factors will most likely also be present to account for soft-gluon effects.

The TMD correlators in~\eqref{shorthand} are the gauge invariant 
non-universal (anti)quark/gluon correlators that contain the appropriate Wilson lines for the particular color-flow diagram $D$ that 
contributes to the partonic subprocess $ab{\rightarrow}cd$.
This is the reason why in expression~\eqref{shorthand} the summation 
over cut diagrams $D$ is displayed explicitly.
The hard functions, \emph{i.e.}\ the expressions $\hat\Sigma^{[D]}$ 
of the individual Feynman diagrams are, themselves, not gauge invariant.
For azimuthal dependence originating from only one of the partons one can effectively use the correlators calculated in Ref.~\cite{Bomhof:2006dp}.
Furthermore, in the tree-level discussion employed here the Wilson lines are along the lightlike $n$-direction,
though a non-lightlike $n^2\,{\neq}\,0$ direction may be required when higher-order corrections are taken into account~\cite{Ji:2004wu,Ji:2004xq}.

From momentum conservation on the partonic level it will follow that,
depending on the process, some components of the partonic momenta 
can be measured
(\emph{e.g.}\ in a way similar to the identification of the incoming 
parton momentum  fraction $x$ with the Bjorken scaling variable $x_B$ 
in deep inelastic scattering). This also works for the transverse momenta. 
For instance, for a hadronic scattering process with a two-to-two hard 
subprocess %$a(p_1)b(p_2){\rightarrow}c(k_1)d(k_2)$ 
the structure in~\eqref{shorthand} will appear with a delta function for momentum conservation enforcing the relation 
$p_1{+}p_2{-}k_1{-}k_2\,{\equiv}\,0$. 
There are ways to measure one or several components of
$q_T\,{\equiv}\,p_{1T}{+}p_{2T}{-}k_{1T}{-}k_{2T}\,
{\approx}\,K_1/z_1{+}K_2/z_2{-}x_1P_1{-}x_2P_2$,
which is not required to vanish by momentum conservation since the directions of the intrinsic transverse momenta of the partons can be different for each observed hadron
(in back-to-back jet production in hadron-hadron scattering it is the component along the outgoing jet direction in the plane perpendicular to the beam axis that is experimentally accessible through the relation 
$q_T{\cdot}\hat K_{\text{jet}}^\perp\,{\propto}\,\sin(\delta\phi)$,
where $\delta\phi$ is the azimuthal imbalance of the two jets in the perpendicular plane~\cite{Boer:2003tx,Bacchetta:2005rm,Bomhof:2006ra,Bomhof:2007su}).
This quantity defines a scale much smaller than the hard scale of the process. 
Using these components one can construct integrated and weighted hadronic cross sections. 
Integrated cross sections will involve the structure
%From momentum conservation on the partonic level it will follow that,
%depending on the process, some components of the partonic momenta 
%can be measured
%(for example in a way similar to the identification of the incoming 
%parton momentum  fraction $x$ with the Bjorken scaling variable $x_B$ 
%in deep inelastic scattering). This also works for the transverse momenta. 
%For instance, for a hard scattering process with a two-to-two hard 
%subprocess there are ways to measure one or several components of
%$q_T\,{\equiv}\,p_{1T}{+}p_{2T}{-}k_{1T}{-}k_{2T}\,
%{\approx}\,K_1/z_1{+}K_2/z_2{-}x_1P_1{-}x_2P_2$.
%This quantity defines a scale much smaller than the hard scale $\sqrt s$ of the %process. 
%Using these components one can construct integrated and weighted hadronic cross %sections. 
%Integrated cross sections will involve the structure
\begin{align}
\Sigma(x_1{,}x_2{,}z_1{,}z_2)
&={\int}d^2p_{1T}\,d^2p_{2T}\,d^2k_{1T}\,d^2k_{2T}\ \;
\Sigma(p_1{,}p_2{,}k_1{,}k_2)\nonumber\\
%%%%
&=\sum_{a{,}{\cdots}{,}d}
\tr\big\{\,\Phi_a(x_1)\Phi_b(x_2)\,\,
\hat\Sigma_{ab\rightarrow cd}(x_1{,}x_2{,}z_1{,}z_2)\,\,
\Delta_c(z_1)\Delta_d(z_2)\,\big\}\ ,\label{PAASHAAS}
\end{align}
while weighted cross sections will involve
(making use of the decomposition in Eq.~\eqref{TransverseMomentGeneral})
\begin{align}
\Sigma_{1\,\partial}^\alpha(x_1{,}x_2{,}z_1{,}z_2)
&={\int}d^2p_{1T}^{\phantom{\alpha}}\,d^2p_{2T}^{\phantom{\alpha}}\,
d^2k_{1T}^{\phantom{\alpha}}\,d^2k_{2T}^{\phantom{\alpha}}\ \;
p_{1T}^{\alpha}\ \;\Sigma(p_1{,}p_2{,}k_1{,}k_2)\displaybreak[0]\nonumber\\
%%%%
&=\sum_{a{,}{\cdots}{,}d}\sum_{D}
\tr\big\{\,\Phi_{a\,\partial}^{[D]\,\alpha}(x_1)\Phi_b(x_2)\,\,
\hat\Sigma^{[D]}(x_1{,}x_2{,}z_1{,}z_2)\,\,
\Delta_c(z_1)\Delta_d(z_2)\,\big\}\displaybreak[0]\nonumber\\
%%%%
&=\sum_{a{,}{\cdots}{,}d}
\big[\,\tr\big\{\,\widetilde\Phi_{a\,\partial}^\alpha(x_1)\Phi_b(x_2)\,\,
\hat\Sigma_{ab\rightarrow cd}(x_1{,}x_2{,}z_1{,}z_2)\,\,
\Delta_c(z_1)\Delta_d(z_2)\,\big\}\nonumber\\
&\mspace{100mu}
+\tr\big\{\,\pi\Phi_{a\,G}^\alpha(x_1{,}x_1)\Phi_b(x_2)\,\,
\hat\Sigma_{[a]b\rightarrow cd}(x_1{,}x_2{,}z_1{,}z_2)\,\,
\Delta_c(z_1)\Delta_d(z_2)\,\big\}\,\big]\ ,\label{SINTERKLAAS}
\end{align}
and similar expressions $\Sigma_{2\,\partial}$, $\Sigma_{1'\,\partial}$
and $\Sigma_{2'\,\partial}$ which are obtained by weighting with 
$p_{2T}$, $k_{1T}$ and $k_{2T}$, respectively.
In these expressions only universal collinear correlators appear. 
In Eqs~\eqref{PAASHAAS} and~\eqref{SINTERKLAAS} we have defined the hard functions
\begin{subequations}\label{HARDfuncts}
\begin{gather}
\hat\Sigma_{ab\rightarrow cd}(x_1{,}x_2{,}z_1{,}z_2)
=\sum\nolimits_D \hat\Sigma^{[D]}(x_1{,}x_2{,}z_1{,}z_2)\ ,
%H^*(x_1{,}x_2{,}z_1{,}z_2)\,H(x_1{,}x_2{,}z_1{,}z_2)\ ,
\label{HARDfunctsA}\\
%%%%
\hat\Sigma_{[a]b\rightarrow cd}(x_1{,}x_2{,}z_1{,}z_2)
=\sum\nolimits_D C_G^{[D]}(a)\ \,\hat\Sigma^{[D]}(x_1{,}x_2{,}z_1{,}z_2)\ .
%H^*(x_1{,}x_2{,}z_1{,}z_2)\,H(x_1{,}x_2{,}z_1{,}z_2)\ 
\label{HARDfunctsB}
\end{gather}
\end{subequations}
The factors $C_G^{[D]}(a)$ are the gluonic pole strengths that appear 
in the decomposition of the transverse moment of the TMD correlator 
of parton $a$.
In expression~\eqref{SINTERKLAAS} this parton was implicitly taken to 
be a quark. If it were a gluon there would have been 
two $\hat\Sigma_{[g]b\rightarrow cd}$ terms,
one corresponding to each of the gluonic pole matrix elements 
%$\Phi_G^{g(f)}$ and $\Phi_G^{g(d)}$, 
$\Gamma_{G_f}$ and $\Gamma_{G_d}$, 
\emph{cf.}\ Eq.~\eqref{GluonDecomp} or Ref.~\cite{Bomhof:2006ra}.
The hard functions in Eqs~\eqref{HARDfunctsA} 
and~\eqref{HARDfunctsB} no longer depend on the individual 
(diagrammatic) contributions $D$, 
but only on the hard process $ab{\rightarrow}cd$.
Moreover, in contrast to the hard functions $\hat\Sigma^{[D]}$ that 
appear in~\eqref{shorthand},
they are gauge invariant expressions.
After performing the traces the $\hat\Sigma_{ab\rightarrow cd}$ reduce 
to the partonic cross sections $d\hat\sigma_{ab\rightarrow cd}$ and the 
$\hat\Sigma_{[a]b\rightarrow cd}$ reduce to the gluonic pole cross sections 
$d\hat\sigma_{[a]b\rightarrow cd}$ calculated 
in Refs~\cite{Bacchetta:2005rm,Bomhof:2006ra,Bacchetta:2007sz}.

\section{Transverse Momentum Dependent Correlators\label{TMDSection}}

To study azimuthal asymmetries arising from one of the partons in hadronic processes mediated by $2{\rightarrow}2$ partonic subprocesses at tree-level it is possible, as we will show explicitly, to organize the TMD correlators in a decomposition analogous to~\eqref{TransverseMomentGeneral} containing TMD correlators with special properties:
\begin{equation}\label{DECOMPOSITION}
\Phi^{[D]}(x{,}p_T)
=\Phi^{(ab\rightarrow cd)}(x{,}p_T)
+C_G^{[D]}\,\pi\Phi_G^{(ab\rightarrow cd)}(x{,}p_T)\ .
\end{equation}
Here $D$ refers to a particular cut Feynman diagram that contributes to the 
cross section of the partonic process $ab{\rightarrow}cd$. 
The gluonic pole factors $C_G^{[D]}$ are the same as those in the 
decomposition of the collinear correlators in~\eqref{TransverseMomentGeneral}.
An important difference between the decomposition of the collinear correlator in~\eqref{TransverseMomentGeneral} and the decomposition of the TMD correlator in~\eqref{DECOMPOSITION} is that the matrix elements in the latter decomposition are not universal in general.
They depend on the partonic process $ab{\rightarrow}cd$ but, 
in contrast to the TMD quark correlators $\Phi^{[D]}$ on the l.h.s.\ of the decomposition,
they do not depend on the individual cut Feynman diagram $D$.
The only diagram dependence resides in the gluonic pole factors.
The matrix elements on the r.h.s.\ of~\eqref{DECOMPOSITION} have been chosen such that they reduce to the familiar universal (process independent) collinear matrix elements when integrating over or weighting with the intrinsic transverse momenta:
\begin{subequations}\label{INTEGRATION}
\begin{alignat}{2}
&{\int}d^2p_T\ \,\Phi^{(ab\rightarrow cd)}(x{,}p_T)
=\Phi(x)\ ,&\qquad
%%%%
&{\int}d^2p_T^{\phantom{\alpha}}\ \,p_T^\alpha\ \,
\Phi^{(ab\rightarrow cd)}(x{,}p_T)
=\widetilde\Phi_\partial^\alpha(x)\ ,\\
%%%%
%%%%
&{\int}d^2p_T\ \,\Phi{}_G^{(ab\rightarrow cd)}(x{,}p_T)=0\ ,&\qquad
%%%%
&{\int}d^2p_T^{\phantom{\alpha}}\ \,p_T^\alpha\ \,
\Phi{}_G^{(ab\rightarrow cd)}(x{,}p_T)
=\Phi_G^\alpha(x{,}x)\ .
\end{alignat}
\end{subequations}

The most straightforward illustration of the decomposition~\eqref{DECOMPOSITION} for quark TMDs are the quark correlators $\Phi^{[+]}(x{,}p_T)$ in SIDIS and $\Phi^{[-]}(x{,}p_T)$ in Drell-Yan
scattering, which contain the simple future and past-pointing Wilson lines
$\mathcal U^{[+]}$ and $\mathcal U^{[-]}$, respectively.
For those correlators one has~\cite{Boer:2003cm}
\begin{equation}
\Phi^{[\pm]}(x{,}p_T)
=\Phi^{(T-\text{even})}(x{,}p_T)+C_G^{[\pm]}\,\Phi^{(T-\text{odd})}(x{,}p_T)\ ,
\label{STANDARD}
\end{equation}
with the $T$-even and $T$-odd quark correlators
\begin{subequations}
\begin{gather}
\Phi^{(T\text{-even})}(x{,}p_T)
=\tfrac{1}{2}\big(\,\Phi^{[+]}(x{,}p_T)\,{+}\,\Phi^{[-]}(x{,}p_T)\,\big)\ ,
\label{DEFINITE-1}\\
%%%%
\Phi^{(T\text{-odd})}(x{,}p_T)
=\tfrac{1}{2}\big(\,\Phi^{[+]}(x{,}p_T)\,{-}\,\Phi^{[-]}(x{,}p_T)\,\big)\ .
\label{DEFINITE-2}
\end{gather}
\end{subequations}
The factors $C_G^{[\pm]}\,{=}\,{\pm}1$ are the same as those for the transverse moments~\eqref{TransverseMomentGeneral} in those processes.

In contrast to~\eqref{STANDARD}, we observe that the TMD matrix elements 
$\Phi^{(ab\rightarrow cd)}(x{,}p_T)$ 
and $\pi\Phi_G^{(ab\rightarrow cd)}(x{,}p_T)$ on the r.h.s.\ of~\eqref{DECOMPOSITION} in general do not have definite behavior under time-reversal.
However, the process dependent universality-breaking parts of the TMD 
correlators can be separated from the universal $T$-even and $T$-odd parts 
in~\eqref{DEFINITE-1} and \eqref{DEFINITE-2}:
\begin{subequations}\label{DeComp}
\begin{gather}
\Phi^{(ab\rightarrow cd)}(x{,}p_T)
=\Phi^{(T\text{-even})}(x{,}p_T)
+\delta\Phi^{(ab\rightarrow cd)}(x{,}p_T)\ ,\\
%%%%
\pi\Phi_G^{(ab\rightarrow cd)}(x{,}p_T)
=\Phi^{(T\text{-odd})}(x{,}p_T)
+\pi\delta\Phi{}_G^{(ab\rightarrow cd)}(x{,}p_T)\ .
\end{gather}
\end{subequations}
In these expressions all process dependence due to Wilson lines on the 
light-front is now contained in process-dependent
universality-breaking matrix elements
$\delta\Phi^{(ab\rightarrow cd)}(x{,}p_T)$ and 
$\pi\delta\Phi{}_G^{(ab\rightarrow cd)}(x{,}p_T)$, which we will
refer to as {\em junk-TMD}.
Also these in general have no definite behavior under time-reversal,
but they do have the special properties that they vanish after $p_T$-integration and weighting:
\begin{subequations}\label{special}
\begin{align}
{\int}d^2p_T\ \,\delta\Phi^{(ab\rightarrow cd)}(x{,}p_T)
&={\int}d^2p_T^{\phantom{\alpha}}\ \,p_T^\alpha\ \,
\delta\Phi^{(ab\rightarrow cd)}(x{,}p_T)=0\ ,\\
%%%%
{\int}d^2p_T\ \,\delta\Phi{}_G^{(ab\rightarrow cd)}(x{,}p_T)
&={\int}d^2p_T^{\phantom{\alpha}}\ \,p_T^\alpha\ \,
\delta\Phi{}_G^{(ab\rightarrow cd)}(x{,}p_T)=0\ .
\end{align}
\end{subequations}
This is consistent with the properties expressed in~\eqref{INTEGRATION}. 
The expressions in Eq.~\eqref{special} will be used in section~\ref{Beleg} to show that the universality-breaking matrix elements vanish in integrated and weighted hadronic cross sections.
Moreover, as can be seen from their explicit expressions in appendix~\ref{CORRS} the (anti)quark/gluon universality-breaking matrix elements vanish in an order $g$ expansion of the Wilson lines 
(\emph{i.e.}\ the one-gluon radiation contribution).

All universality-breaking matrix elements that occur at tree-level in 
proton-proton scattering with hadronic final 
states are listed in appendix~\ref{CORRS}.
The TMD correlators $\Phi^{[D]}(x{,}p_T)$ in these processes have already 
been derived in Ref.~\cite{Bomhof:2006dp} and are given in the tables of 
that reference
(the TMD correlators and gluonic pole factors that appear at tree-level 
in direct photon-jet production in proton-proton scattering can be found 
in Ref.~\cite{Bomhof2007}).
It is straightforward to verify that these results are reproduced with 
the matrix elements given in appendix~\ref{CORRS} and through the 
decompositions~\eqref{DECOMPOSITION} and~\eqref{DeComp}.
This should not come as a surprise, since the matrix elements in the appendix were defined that way.
It is a remarkable and non-trivial observation that with the gluonic pole strengths all the quark correlators encountered in a certain partonic process $ab{\rightarrow}cd$ can be decomposed in terms of only the two matrix elements $\Phi^{(ab\rightarrow cd)}$ and $\pi\Phi^{(ab\rightarrow cd)}_G$ 
(with the properties in Eq.~\eqref{INTEGRATION}).
It should be mentioned, though, that this decomposition is not unique.
For instance, one could also have made a decomposition in terms of more matrix elements.
That is,
by including matrix elements that do not contribute to the zeroth 
($p_T$-integration) and first ($p_T$-weighting) transverse moments in $p_T$,
but do contribute to the second moment, third moment, etc.
It is conceivable that the inclusion of these additional matrix elements will allow one to summarize the TMD quark correlators encountered in different partonic processes, 
in the same way as it was possible to summarize all quark correlators associated to the different Feynman diagrams $D$ that contribute to one specific partonic process $ab{\rightarrow}cd$ by the two matrix elements $\Phi^{(ab\rightarrow cd)}$ and $\pi\Phi^{(ab\rightarrow cd)}_G$.
At present this is just speculation, though,
and the verification or falsification will require more insight into the way that the different Wilson line structures contribute to higher transverse moments.
However, regardless of all these cautionary remarks we believe that the notational advantage, 
the points concerning gauge invariance of the hard functions that will be addressed in section~\ref{Beleg} and the possible role that it could play in relating the gauge link formalism to the results of Refs~\cite{Qiu:2007ar,Qiu:2007ey}
provide more than enough justification for the decomposition in Eq.~\eqref{DECOMPOSITION}.

In the case of gluon distributions we start by defining $T$-even and $T$-odd gluon correlators (\emph{cf}.\ Figs~\ref{simplelinks}c-f),
\begin{subequations}
\begin{gather}
\Gamma^{(T\text{-even})}(x{,}p_T)
=\tfrac{1}{2}\big(\,\Gamma^{[+,+^\dagger]}(x{,}p_T)\,{+}\,
\Gamma^{[-,-^\dagger]}(x{,}p_T)\,\big)\ ,
%={\int}\frac{d(\xi{\cdot}P)d^2\xi_T}{(2\pi)^3}\ e^{ip\cdot\xi}
%\langle P{,}S|\,\tr\big[\,\tfrac{1}{2}F(0)\,\mathcal U^{[+]}\,F(\xi)\,
%\mathcal U^{[+]\,\dagger}
%+\tfrac{1}{2}F(0)\,\mathcal U^{[-]}\,F(\xi)\,
%\mathcal U^{[-]\,\dagger}\,\big]\,|P{,}S\rangle\,\big\rfloor_{\text{LF}}\ ,
\displaybreak[0]\\
%%%%
\Gamma_{(f)}^{(T\text{-odd})}(x{,}p_T)
=\tfrac{1}{2}\big(\,\Gamma^{[+,+^\dagger]}(x{,}p_T)\,{-}\,
\Gamma^{[-,-^\dagger]}(x{,}p_T)\,\big)\ ,
%={\int}\frac{d(\xi{\cdot}P)d^2\xi_T}{(2\pi)^3}\ e^{ip\cdot\xi}
%\langle P{,}S|\,\tr\big[\,\tfrac{1}{2}F(0)\,\mathcal U^{[+]}\,F(\xi)\,
%\mathcal U^{[+]\,\dagger}
%-\tfrac{1}{2}F(0)\,\mathcal U^{[-]}\,F(\xi)\,
%\mathcal U^{[-]\,\dagger}\,\big]\,|P{,}S\rangle\,\big\rfloor_{\text{LF}}\ ,
\displaybreak[0]\\
%%%%
\Gamma_{(d)}^{(T\text{-odd})}(x{,}p_T)
=\tfrac{1}{2}\big(\,\Gamma^{[+,-^\dagger]}(x{,}p_T)\,{-}\,
\Gamma^{[-,+^\dagger]}(x{,}p_T)\,\big)\ ,
%={\int}\frac{d(\xi{\cdot}P)d^2\xi_T}{(2\pi)^3}\ e^{ip\cdot\xi}
%\langle P{,}S|\,\tr\big[\,
%\tfrac{1}{2}F(0)\,\mathcal U^{[+]}\,F(\xi)\,\mathcal U^{[-]\,\dagger}
%-\tfrac{1}{2}F(0)\,\mathcal U^{[-]}\,F(\xi)\,\mathcal U^{[+]\,\dagger}\,\big]\,
%|P{,}S\rangle\,\big\rfloor_{\text{LF}}\ ,
\end{gather}
\end{subequations}
where as in Ref.~\cite{Boer:2003cm} a gluon correlator is called $T$-odd if it vanishes when identifying the future and past-pointing Wilson lines. 
Note that in contrast to $\Gamma^{(T\text{-even})}$ and 
$\Gamma_{(f)}^{(T\text{-odd})}$, 
the correlator $\Gamma_{(d)}^{(T\text{-odd})}$ cannot be written as a 
matrix element of two gluon fields with a single Wilson line in the adjoint 
representation.  
After $p_T$-integration the $T$-even and $T$-odd correlators reduce to the universal collinear gluon matrix elements in the expressions in Eqs~\eqref{Gluon} and~\eqref{GluonDecomp}: 
\begin{subequations}
\begin{alignat}{2}
&{\int}d^2p_T\ \;\Gamma^{(T\text{-even})}(x{,}p_T)
=\Gamma(x)\ ,&\qquad
&{\int}d^2p_T^{\phantom{\alpha}}\ \;p_T^\alpha\ \;
\Gamma^{(T\text{-even})}(x{,}p_T)
=\widetilde\Gamma_\partial^\alpha(x)\ ,\\
%%%%
&{\int}d^2p_T\ \;\Gamma_{(f)}^{(T\text{-odd})}(x{,}p_T)=0\ ,&
&{\int}d^2p_T^{\phantom{\alpha}}\ \;p_T^\alpha\ \;
\Gamma_{(f)}^{(T\text{-odd})}(x{,}p_T)
=\pi\Gamma_{G_f}^\alpha(x{,}x)\ ,\\
%%%%
&{\int}d^2p_T\ \;\Gamma_{(d)}^{(T\text{-odd})}(x{,}p_T)=0\ ,&
&{\int}d^2p_T^{\phantom{\alpha}}\ \;p_T^\alpha\ \;
\Gamma_{(d)}^{(T\text{-odd})}(x{,}p_T)
=\pi\Gamma_{G_d}^\alpha(x{,}x)\ .
\end{alignat}
\end{subequations}
Since there are two distinct ways to construct $T$-odd 
gluon correlators, it will follow in the next section that there are also two distinct TMD gluon-Sivers distribution functions 
(\emph{cf.} Eq.~\eqref{GluonCorrB}).
There is actually also a second way to construct a $T$-even gluon correlator:
$\Gamma^{\prime\,(T\text{-even})}\,
{=}\,\frac{1}{2}(\Gamma^{[+,-^\dagger]}\,{+}\,\Gamma^{[-,+^\dagger]})$.
However, this correlator is not needed in the 
decomposition~\eqref{DECOMPOSITION2} of the TMD gluon correlators,
since the difference between $\Gamma^{\prime\,(T\text{-even})}$ and 
$\Gamma^{(T\text{-even})}$ is a matrix element that vanishes upon 
$p_T$-integration and $p_T$-weighting.
This difference may, therefore, be absorbed in the universality-breaking 
matrix elements $\delta\Gamma^{(ab\rightarrow cd)}$ to be defined 
in~\eqref{Pizza1}.

A decomposition resembling the one in~\eqref{GluonDecomp} for TMD gluon correlators can almost be made:
\begin{equation}\begin{split}\label{DECOMPOSITION2}
\Gamma^{[D]}(x{,}p_T)
=\Gamma^{(ab\rightarrow cd)}(x{,}p_T)
+C_G^{(f)\,[D]}\,\Gamma_{G_f}^{(ab\rightarrow cd)}(x{,}p_T)
+C_G^{(d)\,[D]}\,\Gamma_{G_d}^{(ab\rightarrow cd)}(x{,}p_T)\ .
\end{split}\end{equation}
This leaves only a specific type of matrix
elements with colorless intermediate states unaccounted for 
(we will return to this point in a moment). 
The TMD matrix elements on the r.h.s.\ of~\eqref{DECOMPOSITION2} only depend 
on the process and not on the particular Feynman diagram $D$ that 
contributes to that process.
The multiplicative factors $C_G^{(f)}$ and $C_G^{(d)}$ are 
the gluonic pole factors calculated in Ref.~\cite{Bomhof:2006ra},
the same that also appear in the decomposition~\eqref{GluonDecomp} of the collinear correlator.
Under $p_T$-integration and weighting the matrix elements 
$\Gamma^{(ab\rightarrow cd)}(x{,}p_T)$ and 
$\Gamma_{G_{f/d}}^{(ab\rightarrow cd)}(x{,}p_T)$ have the 
same behavior as $\Gamma^{(T\text{-even})}(x{,}p_T)$ and
$\Gamma_{(f/d)}^{(T\text{-odd})}(x{,}p_T)$, respectively.
Therefore, one can make the further separation
\begin{subequations}\label{Pizza}
\begin{gather}
\Gamma^{(ab\rightarrow cd)}(x{,}p_T)
=\Gamma^{(T\text{-even})}(x{,}p_T)
+\delta\Gamma^{(ab\rightarrow cd)}(x{,}p_T)\ ,\label{Pizza1}\\
%%%%
\pi\Gamma_{G_{f/d}}^{(ab\rightarrow cd)}(x{,}p_T)
=\Gamma_{(f/d)}^{(T\text{-odd})}(x{,}p_T)
+\pi\delta\Gamma_{G_{f/d}}^{(ab\rightarrow cd)}(x{,}p_T)\ ,\label{Pizza2}
\end{gather}
\end{subequations}
in which all process-dependence due to Wilson lines on the light-front has been 
gathered in the universality-breaking 
matrix elements $\delta\Gamma^{(ab\rightarrow cd)}(x{,}p_T)$ and
$\pi\delta\Gamma_{(f/d)}^{(ab\rightarrow cd)}(x{,}p_T)$, 
which have the special properties that they vanish after a $p_T$-integration 
and $p_T$-weighting.

It is also straightforward to check that through the decompositions in~\eqref{DECOMPOSITION2}-\eqref{Pizza} and with the 
universality-breaking matrix elements given in appendix~\ref{CORRS} the 
TMD gluon correlators in the tables~4,~5 and~8 of Ref.~\cite{Bomhof:2006dp} 
corresponding to $qg{\rightarrow}qg$, 
$\bar qg{\rightarrow}\bar qg$ and $gg{\rightarrow}gg$ scattering are 
reproduced. 
However, in the TMD correlators in the tables~6 and~7 for the 
processes $q\bar q{\rightarrow}gg$ and $gg{\rightarrow}q\bar q$ one
does not recover terms of the form 
$\langle P{,}S|\,\tr\big[F(\xi)\mathcal U^{[\Box]}\big]
\tr\big[F(0)\mathcal U^{[\Box]\,\dagger}\big]\,|P{,}S\rangle$, where
$\mathcal U^{[\Box]}\,{=}\,\mathcal U^{[+]}\mathcal U^{[-]\,\dagger}$.
These matrix elements involve colorless intermediate states
and they do not contribute to the $p_T$-integrated gluon 
correlators $\Gamma(x)$ nor to the first transverse moments 
$\Gamma_\partial^{[D]}(x)$. 
They can be included in~\eqref{DECOMPOSITION2} by adding diagram-dependent universality-breaking matrix elements which will not appear in integrated and weighted hadronic cross sections.
However, it could be that they contribute to the second or higher transverse moments.

\section{Parametrizations of parton correlators\label{Parameterizations}}

\begin{figure}
\centering
\psfrag{pt}[lc][lc]{$|\boldsymbol p_T|^2$}
\includegraphics[width=4cm]{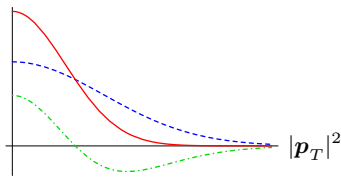}\\
\parbox{0.75\textwidth}{\caption{
Possible behavior of the universal distribution function $f_1(x{,}p_T^2)$ (dashed line) and the process-dependent function 
$f_1^{(ab\rightarrow cd)}(x{,}p_T^2)$ (solid line) as a function of 
$|\boldsymbol p_T|^2$.
Their difference-function, 
$\delta f_1^{(ab\rightarrow cd)}(x{,}p_T^2)$ (dash-dotted line), 
vanishes upon integration over $p_T$.\label{PLOT}}}
\end{figure}

At leading twist the parametrizations of the different TMD quark correlators 
are given by
\begin{subequations}\label{TMDparametrizations}
\begin{align}
\Phi^{(T\text{-even})}(x{,}p_T)
=\tfrac{1}{2}\,
\Big\{\,
&f_1(x{,}p_T^2)\;\slash P
+\tfrac{1}{2}\,h_{1T}(x{,}p_T^2)\;\gamma_5[\slash S_T,\slash P]
\displaybreak[0]\nonumber\\
%%%%
&\ +S_L\;h_{1L}^\perp(x{,}p_T^2)\;
\gamma_5\frac{[\;\slash p_T,\slash P]}{2M}
+\frac{\boldsymbol p_T{\cdot}\boldsymbol  S_T}{M}\;h_{1T}^\perp(x{,}p_T^2)\;
\gamma_5\frac{[\;\slash p_T,\slash P]}{2M}\displaybreak[0]\nonumber\\
%%%%
&\ +S_L\;g_{1L}(x{,}p_T^2)\;\gamma_5\slash P
+\frac{\boldsymbol p_T{\cdot}\boldsymbol  S_T}{M}\;
g_{1T}(x{,}p_T^2)\;\gamma_5\slash P\,\Big\}\ ,\label{TMDparametrizationA}\\
%%%%
%%%%
\Phi^{(T\text{-odd})}(x{,}p_T)
=\tfrac{1}{2}\,
\Big\{\,&ih_1^\perp(x{,}p_T^2)\;\frac{[\;\slash p_T,\slash P]}{2M}
-\frac{\epsilon_T^{p_T S_T}}{M}\;f_{1T}^\perp(x{,}p_T^2)\;\slash P\,
\Big\}\ ,\label{TMDparametrizationB}
\end{align}
and
\begin{align}
\delta\Phi^{(ab\rightarrow cd)}(x{,}p_T)
=\tfrac{1}{2}\,
\Big\{\,&\delta f_1^{(ab\rightarrow cd)}(x{,}p_T^2)\;\slash P
+\tfrac{1}{2}\,\delta h_{1T}^{(ab\rightarrow cd)}(x{,}p_T^2)\;
\gamma_5[\slash S_T,\slash P]
\displaybreak[0]\nonumber\\
%%%%
&\ +S_L\;\delta h_{1L}^{\perp\,(ab\rightarrow cd)}(x{,}p_T^2)\;
\gamma_5\frac{[\;\slash p_T,\slash P]}{2M}
+\frac{\boldsymbol p_T{\cdot}\boldsymbol  S_T}{M}\;
\delta h_{1T}^{\perp\,(ab\rightarrow cd)}(x{,}p_T^2)\;
\gamma_5\frac{[\;\slash p_T,\slash P]}{2M}\displaybreak[0]\nonumber\\
%%%%
&\ +S_L\;\delta g_{1L}^{(ab\rightarrow cd)}(x{,}p_T^2)\;\gamma_5\slash P
+\frac{\boldsymbol p_T{\cdot}\boldsymbol  S_T}{M}\;
\delta g_{1T}^{(ab\rightarrow cd)}(x{,}p_T^2)\;
\gamma_5\slash P\displaybreak[0]\nonumber\\
%%%%
&\ +i\,\delta h_1^{\perp\,(ab\rightarrow cd)}(x{,}p_T^2)\;
\frac{[\;\slash p_T,\slash P]}{2M}
-\frac{\epsilon_T^{p_T S_T}}{M}\;
\delta f_{1T}^{\perp\,(ab\rightarrow cd)}(x{,}p_T^2)\;
\slash P\,\Big\}\ ,\label{TMDparametrizationC}
\end{align}
\end{subequations}
with similar parametrizations for the matrix elements 
$\pi\delta\Phi_G^{(ab\rightarrow cd)}(x{,}p_T)$
containing a (different) set of distribution functions 
$\delta f_1^{([a]b\rightarrow cd)}$, etc.
The quark distribution functions that appear in the 
parametrizations~\eqref{TMDparametrizationA} and~\eqref{TMDparametrizationB} 
are the familiar $T$-even and $T$-odd (respectively) quark distribution functions as measured in SIDIS.
On the other hand, the quark distribution functions in~\eqref{TMDparametrizationC} and in the parametrization of $\delta\Phi_G$ are process dependent. 
From the properties in~\eqref{special} one finds that the functions
$\delta f_1$, $\delta h_{1T}$ and $\delta g_{1L}$ vanish upon $p_T$-integration,
for instance
\begin{equation}
{\int}d^2p_T\ \;\delta f_1^{(ab\rightarrow cd)}(x{,}p_T^2)
=0\ ,\label{pasta1}
\end{equation}
illustrated in Figure~\ref{PLOT}.
Also the functions 
$\delta h_{1L}^{\perp (1)}$, $\delta h_{1T}^{\perp(1)}$,
$\delta g_{1T}^{(1)}$, $\delta h_1^{\perp(1)}$ and $\delta f_{1T}^{\perp(1)}$
vanish, \emph{e.g.},
\begin{equation}
{\int}d^2p_T\ \;\frac{\boldsymbol p_T^2}{2M^2}\,\;
\delta f_{1T}^{\perp\,(ab\rightarrow cd)}(x{,}p_T^2)
=0\ .\label{pasta2}
\end{equation}
The same holds for the corresponding functions in the parametrization of 
$\pi\delta\Phi_G^{(ab\rightarrow cd)}(x{,}p_T)$.

For gluon distribution correlators we use the parameterizations of Ref.~\cite{Mulders:2000sh} with the naming convention discussed in Ref.~\cite{Meissner:2007rx}:
\begin{subequations}\label{GluonCorr}
\begin{gather}
\Gamma^{(T\text{-even})\,\mu\nu}(x{,}p_T)
=\frac{1}{2x}\,\bigg\{-g_T^{\mu\nu}\,f_1^g(x{,}p_T^2)
+\bigg(\frac{p_T^\mu p_T^\nu}{M^2}\,
{+}\,g_T^{\mu\nu}\frac{\boldsymbol p_T^2}{2M^2}\bigg)\;h_1^{\perp g}(x{,}p_T^2)
\nonumber\\
%%%%
\mspace{250mu}
+i\epsilon_T^{\mu\nu}S_L\;g_{1L}^g(x{,}p_T^2)
+i\epsilon_T^{\mu\nu}\,\frac{\boldsymbol p_T{\cdot}\boldsymbol S_T}{M}\;
g_{1T}^g(x{,}p_T^2)\,\bigg\}\ ,\label{GluonCorrA}\\
%%%%
%%%%
\Gamma_{(f/d)}^{(T\text{-odd})\,\mu\nu}(x{,}p_T)
=\frac{1}{2x}\,\bigg\{\,
g_T^{\mu\nu}\frac{\epsilon_T^{p_TS_T}}{M}\,f_{1T}^{\perp g\,(f/d)}(x{,}p_T^2)
-\frac{\epsilon_T^{p_T\{\mu}S_T^{\nu\}}
{+}\epsilon_T^{S_T\{\mu}p_T^{\nu\}}}{4M}\;
h_{1T}^{g\,(f/d)}(x{,}p_T^2)\nonumber\\
\mspace{250mu}
-\frac{\epsilon_T^{p_T\{\mu}p_T^{\nu\}}}{2M^2}S_L\;
h_{1L}^{\perp g\,(f/d)}(x{,}p_T^2)
-\frac{\epsilon_T^{p_T\{\mu}p_T^{\nu\}}}{2M^2}
\frac{\boldsymbol p_T{\cdot}\boldsymbol S_T}{M}\;
h_{1T}^{\perp g\,(f/d)}(x{,}p_T^2)\,\bigg\}\ .\label{GluonCorrB}
\end{gather}
\end{subequations}
In particular one has two distinct gluon-Sivers distribution functions $f_{1T}^{\perp\,g\,(f)}(x{,}p_T^2)$ and $f_{1T}^{\perp\,g\,(d)}(x{,}p_T^2)$ corresponding to the two ways to construct $T$-odd gluon correlators.
Their first transverse moments are the functions $G_T^{(f/d)}{}^{(1)}(x)\,{\equiv}\,f_{1T}^{\perp\,g\,(f/d)}{}^{(1)}(x)$ introduced in Ref.~\cite{Bomhof:2006ra}.
The matrix elements $\delta\Gamma^{(ab\rightarrow cd)}(x{,}p_T^2)$ and 
$\pi\delta\Gamma_{G_{f/d}}^{(ab\rightarrow cd)}(x{,}p_T^2)$
contain both $T$-even and $T$-odd gluon distribution functions.
These functions, however, will vanish under $p_T$-integration or weighting,
as for the quark universality-breaking distribution functions
(\emph{cf}.~\eqref{pasta1} and~\eqref{pasta2}).

The analysis above can be extended to antiquark distribution correlators in the obvious way. Also the treatment of (anti)quark fragmentation correlators is 
straightforward, with the understanding that in the matrix elements 
$\Delta^{(T\text{-even/odd})}(z{,}k_T)$ the superscripts $T$-even and $T$-odd 
refer to the operator structure in the correlators, 
each of which contain both $T$-even and $T$-odd fragmentation functions
(this also holds for the gluon fragmentation correlators~$\hat\Gamma$).

\section{Transverse momentum dependent functions in hadronic cross sections\label{Beleg}}

The advantage of the decomposition of the TMD quark (and gluon) correlators 
in expression~\eqref{DECOMPOSITION} (and~\eqref{DECOMPOSITION2}) 
is evident: instead of having to list the TMD correlators $\Phi^{[D]}$
($\Gamma^{[D]}$) for every cut Feynman diagram of a process, 
the two (or three in the case of gluons) matrix elements on the r.h.s.\ of 
those decompositions suffice for that particular process.
This observation leads to a reduction in the number of TMD matrix elements 
that need to be considered.
From a bookkeeping point of view this makes the treatment of TMD 
correlators more manageable.
What is perhaps more important is that the decompositions~\eqref{DECOMPOSITION} and~\eqref{DECOMPOSITION2} will allow us to show that also unweighted,
unintegrated hadronic cross sections can be written as products of soft parton correlators and hard partonic functions that are separately manifestly gauge invariant,
analogous to the collinear case in expressions~\eqref{PAASHAAS} and~\eqref{SINTERKLAAS}.
Moreover, the hard partonic functions are the partonic cross sections or the gluonic pole cross sections.
This is what will be argued in this section.
The proper context should be within a transverse momentum dependent factorization theorem.
For most processes such a theorem does not exist yet.
We will therefore take as a starting point the assumption that the hadronic cross sections will factorize in a hard partonic function and a soft parton correlator for each observed hadron separately,
and that gluon initial and final-state interactions will lead to the required Wilson lines.
We believe that this assumption is sufficiently generic for our conclusions to be applicable for many hadronic processes,
in particular to back-to-back dijet or photon-jet production in proton-proton scattering.

By inserting the decompositions~\eqref{DECOMPOSITION} and~\eqref{DECOMPOSITION2} of the TMD parton correlators into the expression for the unintegrated hadronic cross section in Eq.~\eqref{shorthand} the parton contribution to a hard $2{\rightarrow}2$ process becomes
\begin{gather}
\mspace{-10mu}
\Sigma(p_1{,}p_2{,}k_1{,}k_2)\nonumber\\
\mspace{5mu}
=\sum_{a,{\cdots},d}\big\{\,\tr\big[\,
\Phi_a^{(ab{\rightarrow}cd)}(x_1{,}p_{1T})\,
\Phi_b^{(ab{\rightarrow}cd)}(x_2{,}p_{2T})\,\,
\hat\Sigma_{ab{\rightarrow}cd}\,\,
\Delta_c^{(ab{\rightarrow}cd)}(z_1{,}k_{1T})\,
\Delta_d^{(ab{\rightarrow}cd)}(z_2{,}k_{2T})\,\big]\nonumber\\[-2mm]
\mspace{100mu}
+\tr\big[\,\pi\Phi_{a\,G}^{(ab{\rightarrow}cd)}(x_1{,}p_{1T})\,
\Phi_b^{(ab{\rightarrow}cd)}(x_2{,}p_{2T})\,\,
\hat\Sigma_{[a]b{\rightarrow}cd}\,\,
\Delta_c^{(ab{\rightarrow}cd)}(z_1{,}k_{1T})\,
\Delta_d^{(ab{\rightarrow}cd)}(z_2{,}k_{2T})\,\big]\nonumber\\[1mm]
\mspace{100mu}
+\cdots\,\big\}\ ,\label{WALNOOT}
\end{gather}
which forms the central result of this paper.
Again it is implicitly implied that parton $a$ is an (anti)quark,
since if it were a gluon there would be two gluonic pole terms 
$\Gamma_{G_f}\hat\Sigma_{[g]b\rightarrow cd}^{(f)}$ and 
$\Gamma_{G_d}\hat\Sigma_{[g]b\rightarrow cd}^{(d)}$.
In~\eqref{WALNOOT} both terms in the decomposition of the TMD correlator of (anti)quark $a$ have been given explicitly,
while only the first terms of the decompositions~\eqref{DECOMPOSITION} and~\eqref{DECOMPOSITION2} were used for the other partons.
The ``$+\cdots$'' contains the other possible combinations of the 
terms in those decompositions and also the contributions where parton $a$ is a gluon.

The hard functions in expression~\eqref{WALNOOT} are the partonic and gluonic pole cross sections in~\eqref{HARDfunctsA}
and~\eqref{HARDfunctsB}
(in the collinear expansions of the hard functions, 
which have corrections at order $\mathcal O(1/s)$).
Hence, the expression in~\eqref{WALNOOT} demonstrates that by using the decompositions in~\eqref{DECOMPOSITION} and~\eqref{DECOMPOSITION2} and by introducing the gluonic pole cross sections,
also the unintegrated, unweighted hadronic cross section can be written as a product of soft TMD parton correlators and hard partonic functions that are separately and manifestly gauge invariant.
After performing the traces these hard functions reduce to the partonic and gluonic pole cross sections calculated in Refs~\cite{Bacchetta:2005rm,Bomhof:2006ra,Bacchetta:2007sz}.
Hence, it follows that the gluonic pole cross sections that have been seen~\cite{Bacchetta:2005rm,Bomhof:2006ra,Bomhof:2007su,Bacchetta:2007sz} to represent the hard partonic functions in weighted spin asymmetries already appear in the fully TMD cross sections.
This conclusion is consistent with the work in 
Refs~\cite{Qiu:2007ar,Qiu:2007ey}, where a TMD factorization formula 
for the quark-Sivers contribution to single transverse-spin asymmetries 
in dijet production is proposed with the gluonic pole cross sections 
of Refs~\cite{Bacchetta:2005rm,Bomhof:2006ra} as hard functions.
However, the work in Refs~\cite{Qiu:2007ar,Qiu:2007ey} limits to one-gluon exchange and as a result it obtains the distribution functions 
measured in SIDIS (\emph{i.e.}, 
transverse momentum dependent distribution functions with a future pointing Wilson line in the hadronic matrix elements defining them). 
In contrast, our expression~\eqref{WALNOOT} involves correlators with process dependent Wilson lines in the operator definitions and a factorized form with universal distribution functions would only be achieved if,
regardless of the appearance of process dependent Wilson lines in their definitions,
the TMD matrix elements in~\eqref{WALNOOT} are identical for all partonic channels 
%in dijet production $pp{\rightarrow}jj'X$ 
and equal those in semi-inclusive deep inelastic scattering.
In the present context this translates into a vanishing of all universality-breaking matrix elements,
for which we see no reason.
Indeed, universality has recently also been disputed in Ref.~\cite{Collins:2007nk},
where it is argued that a TMD factorization theorem for this process with universal distribution functions is not possible.
This is confirmed by an explicit calculation including the exchange of two collinear gluons~\cite{Collins:2007jp}.
A recent extension~\cite{Vogelsang:2007jk} of the work in Refs~\cite{Qiu:2007ar,Qiu:2007ey} including two-gluon exchange also points to non-universality.
Moreover, it shows that Refs~\cite{Bomhof:2004aw,Bacchetta:2005rm,Bomhof:2006dp,Bomhof:2006ra},
\cite{Qiu:2007ar,Qiu:2007ey} and~\cite{Collins:2007nk} are consistent to 
(at least) that order.
We want to emphasize that for a full connection the role of soft factors,
which have been neglected in the present study,
should also be investigated.

With the parametrizations~\eqref{TMDparametrizations} 
and~\eqref{GluonCorr} inserted in~\eqref{WALNOOT}, 
an expression for the hadronic cross section in terms of TMD parton 
distribution and fragmentation functions is obtained.
After performing the traces these hard functions reduce to the partonic 
and gluonic pole cross sections encountered in
Refs~\cite{Bacchetta:2005rm,Bomhof:2006ra,Bomhof:2007su,Bacchetta:2007sz}.
As an illustration we consider the contribution to~\eqref{WALNOOT} of an unpolarized quark in an unpolarized hadron, 
important for unintegrated spin averaged hadronic cross sections
(summations over parton types are understood):
%As an illustration we consider the contribution to~\eqref{WALNOOT} of one %unpolarized quark in an unpolarized hadron, 
%important for spin averaged hadronic cross sections
%(limiting to one of the ).
%The unintegrated cross ssection, then, involves 
%(summations over parton types are understood)
\begin{subequations}\label{NotIntegrated}
\begin{equation}
d\sigma_U
\sim\ 
\big\{\,\underbrace{f_1(x_1^{\phantom{1}}{,}p_{1T}^2)
+\delta f_1^{(ab\rightarrow cd)}(x_1^{\phantom{1}}{,}p_{1T}^2)}_{
f_1^{(ab\rightarrow cd)}(x_1^{\phantom{1}}{,}p_{1T}^2)}\,\big\}\,\,
d\hat\sigma_{ab\rightarrow cd}
+\delta f_1^{([a]b\rightarrow cd)}(x_1^{\phantom{1}}{,}p_{1T}^2)\,\,
d\hat\sigma_{[a]b\rightarrow cd}\ .\label{NotIntegratedA}
\end{equation}
Taking as a second example the contribution of an unpolarized quark in a transversally polarized hadron, 
important for unintegrated single-spin asymmetries,
one finds
\begin{equation}
d\sigma_T
\sim\ 
%\frac{\boldsymbol{\hat P}{\cdot}(\boldsymbol S_T{\times}\boldsymbol p_{1T})}
%{M}\,\Big[\,
\delta f_{1T}^{\perp\,(ab\rightarrow cd)}(x_1^{\phantom{1}}{,}p_{1T}^2)\,\,
d\hat\sigma_{ab\rightarrow cd}
+\big\{\,\underbrace{f_{1T}^\perp(x_1^{\phantom{1}}{,}p_{1T}^2)
+\delta f_{1T}^{\perp\,([a]b\rightarrow cd)}(x_1^{\phantom{1}}{,}p_{1T}^2)}_{
f_{1T}^{\perp\,(ab\rightarrow cd)}(x_1^{\phantom{1}}{,}p_{1T}^2)}\,\big\}\,\,
d\hat\sigma_{[a]b\rightarrow cd}
%\,\,\Big]
\ .\label{NotIntegratedB}
\end{equation}
\end{subequations}
The terms with only universal $T$-even functions will appear folded 
with partonic cross sections and the universal $T$-odd functions 
with gluonic pole cross sections.
In addition, there are various universality-breaking functions that appear 
with partonic cross sections or gluonic pole cross sections.
A shorter notation for some of the terms in these expressions could have been obtained by not extracting the universal $T$-even and $T$-odd parts of the distribution functions $f_1$ and $f_{1T}^\perp$, 
as indicated by the underbraces in~\eqref{NotIntegrated}.
In particular, due to these universality-breaking functions the gluonic pole cross sections also appear in the unweighted spin-averaged cross sections~\eqref{NotIntegratedA} and the usual partonic cross sections also appear in the unweighted single-spin asymmetries~\eqref{NotIntegratedB}.
However, in the light of the properties in~\eqref{pasta1}-\eqref{pasta2} it is seen that terms with universality-breaking matrix elements do not contribute to the integrated and $q_T$-weighted 
(where $q_T$ is as defined in section~\ref{Boterham}) 
hadronic cross sections which are expressed in terms of the structures in~\eqref{PAASHAAS} and~\eqref{SINTERKLAAS}:
\begin{subequations}\label{ColLinear}
\begin{alignat}{2}
&\text{integrated:}&\mspace{20mu}&
\langle\,d\sigma\,\rangle
\sim \Sigma(x_1{,}x_2)
\propto f_1(x_1)\,d\hat\sigma_{ab\rightarrow cd}\ ,\\
%%%%
&\text{weighted:}&&
\langle\,q_T\,d\sigma\,\rangle
\sim 
\big(\Sigma_{1\,\partial}+\Sigma_{2\,\partial}\big)(x_1{,}x_2)
\propto f_{1T}^{\perp(1)}(x_1)\,d\hat\sigma_{[a]b\rightarrow cd}\ .
\label{WeiGhted}
\end{alignat}
\end{subequations}
For back-to-back jet production in polarized proton-proton scattering 
($p^\uparrow p{\rightarrow}jjX$) this (in essence) reproduces the results of Refs~\cite{Bacchetta:2005rm,Bomhof:2006ra,Bomhof:2007su},
while for photon-jet production ($p^\uparrow p{\rightarrow}\gamma jX$) it reproduces the results in Ref.~\cite{Bacchetta:2007sz}.

Only if the universality-breaking matrix elements vanish in the unintegrated, unweighted cross sections~\eqref{WALNOOT} and~\eqref{NotIntegrated}
does one also in the TMD case arrive at the situation with universal functions only.
Otherwise, the non-universality of the unweighted processes is important.
It will affect the results of experiments that try to look at explicit $p_T$-dependence or that construct weighted cross sections involving convolutions of TMD functions which upon integration do not factorize into transverse moments, 
\emph{e.g.}\ when looking at $\langle\,\sin(\phi_h{\pm}\phi_S)\,\rangle$, 
rather than $\langle\,P_{\pi\perp}\sin(\phi_h{\pm}\phi_S)\,\rangle$ asymmetries in SIDIS to extract transversity or Sivers functions.

%As a final remark we note that if one wants to extend Eq.~\eqref{WALNOOT} in %order to allow for double or higher $p_T$-weighting, one should realize that %already in the collinear case one will encounter
%additional gluonic pole matrix elements with new color factors.
%With these additional structures one might reduce the process-dependence in the %TMD-decompositions in Eqs.~\eqref{DECOMPOSITION} and \eqref{DECOMPOSITION2} and %have functions for which Eq.~\eqref{INTEGRATION} is
%extended one step further, \emph{i.e.}\ the TMD process-dependent functions
%reduce to specific universal collinear functions after
%integration, single and double weighting.

\section{Summary\label{Conclusion}}

We have argued that the gluonic pole cross sections,
the hard partonic scattering functions that are folded with the collinear 
parton distribution functions in weighted spin-asymmetries,
also appear in unintegrated, unweighted hadronic cross sections.
Assuming as a starting point that the hadronic cross section factorizes at the diagrammatic level in a hard partonic function and for each of the observed hadrons a soft parton correlator, 
we have shown that the transverse momentum dependent 
cross section can be written in terms of soft and hard functions that 
are separately manifestly gauge-invariant.
The hard functions are the partonic cross sections or the gluonic pole 
cross sections.
The latter do not show up in the integrated cross sections.
The soft functions are the TMD parton distribution (and fragmentation) 
functions with Wilson lines on the light-front in their field theoretical operator definitions.  
These Wilson lines can be considered as the 
collective effect of gluon initial and final state interactions.
Since they are process dependent,
the TMD parton distributions are in general non-universal.
By systematically separating the universality-breaking parts from the 
universal $T$-even and $T$-odd parts,
we arrived at an expression that has soft parts multiplying the partonic and gluonic pole cross sections,
as was also found in Refs~\cite{Qiu:2007ar,Qiu:2007ey}.
%up to non-universal terms which vanish in the integrated and weighted 
%cross sections considered in 
%Refs~\cite{Bacchetta:2005rm,Bomhof:2006ra,Bomhof:2007su},
However, in the gauge link approach taken here the gluonic pole cross section can also emerge in TMD spin-averaged processes and that ordinary partonic cross sections can also arise in TMD single-spin asymmetries.
They appear with universality-breaking functions that will vanish for the integrated and weighted processes considered in Refs~\cite{Bacchetta:2005rm,Bomhof:2006ra,Bomhof:2007su,Bacchetta:2007sz}.
They also vanish at the level of one-gluon radiation contributions,
which corresponds to the order $g$ term of the Wilson lines.

The non-universal terms are well-defined matrix elements. 
All universality-breaking matrix elements that are encountered at tree-level 
in proton-proton scattering with $2{\rightarrow}2$ partonic processes 
have been calculated and are given in the appendix of this paper. 
In particular, the process-dependent universality-breaking matrix elements $\delta\Phi$, etc.\ 
disappear in the simple electroweak processes with underlying hard parts like 
$q\gamma^\ast{\rightarrow}q$ and $q\bar q{\rightarrow}\gamma^\ast$. 
We believe that the explicit identification of universality-breaking matrix elements is an important contribution to arrive at a unified picture of TMD factorization of hadronic scattering processes.

\begin{acknowledgments}
We are grateful to Alessandro Bacchetta, Dani\"el Boer, John Collins, 
Umberto D'Alesio, Leonard Gamberg, Andreas Metz, Peter Schweitzer, Werner Vogelsang and 
Feng Yuan for useful discussions.
The work of C.B.\ was supported by the foundation for Fundamental 
Research of Matter (FOM) and the National Organization for Scientific 
Research (NWO).
\end{acknowledgments}

\appendix

\section{Universality-Breaking Matrix Elements\label{CORRS}}

We list the universality-breaking matrix elements that appear at tree-level 
in proton-proton scattering.
To improve readability we employ the schematic notation
$(2\pi)^{-3}{\int}d(\xi{\cdot}P)d^2\xi_T\ e^{ip\cdot\xi}
\langle P{,}S|\,\overline \psi(0)\,\mathcal U\,
\psi(\xi)\,|P{,}S\rangle\,\big\rfloor_{\text{LF}}
\propto
\langle\,\overline \psi(0)\,\mathcal U\,\psi(\xi)\,\rangle$,
and similarly for the other parton correlators.
For the $2{\rightarrow}2$ partonic channels with four colored external legs we only list the quark and/or gluon distribution correlators.
As can be seen from the tables in Ref.~\cite{Bomhof:2006dp} the TMD correlators of the other partons are very similar in structure to the correlators considered here and can straightforwardly be obtained by comparing the results below to the tables of the stated reference.
In particular the correlators in $\bar qg{\rightarrow}\bar qg$ scattering are simply obtained by comparing to those in $qg{\rightarrow}qg$ scattering.
Similarly, the fragmentation correlators in 
$q\bar q{\rightarrow}gg$ ($gg{\rightarrow}q\bar q$) scattering can be obtained by comparing to the distribution correlators in 
$gg{\rightarrow}q\bar q$ ($q\bar q{\rightarrow}gg$).\\[2mm]

\begin{small}
\begin{gather}
\textstyle{\boxed{\ q\gamma^*{\rightarrow}q \quad\  
q\bar q{\rightarrow}\gamma^*\ }}\nonumber\\
%%%%
\delta\Phi^{(q\gamma^*\rightarrow q)}(x{,}p_T)
=\pi\delta\Phi_G^{(q\gamma^*\rightarrow q)}(x{,}p_T)
%=\delta\Phi^{(\gamma^*\rightarrow q\bar q)}(x{,}p_T)
=0\\
%%%%
\delta\Phi^{(q\bar q\rightarrow\gamma^*)}(x{,}p_T)
=\pi\delta\Phi_G^{(q\bar q\rightarrow\gamma^*)}(x{,}p_T)
%=\delta\Phi_G^{(\gamma^*\rightarrow q\bar q)}(x{,}p_T)
=0%\\
%\nonumber
\end{gather}
\end{small}

%\begin{center}
%$\boxed{\ q\bar q{\rightarrow}g\gamma\ }$
%\end{center}
\begin{small}
\begin{gather}
\textstyle{\boxed{\ q\bar q{\rightarrow}g\gamma\ }}\nonumber\\
%%%%
\delta\Phi^{(q\bar q\rightarrow g\gamma)}(x{,}p_T)
=\pi\delta\Phi_G^{(q\bar q\rightarrow g\gamma)}(x{,}p_T)
\propto\langle\,\overline \psi(0)\,
\Big\{\,\tfrac{1}{2}\frac{\tr[\mathcal U^{[\Box]\dagger}]}{N}\mathcal U^{[+]}
-\tfrac{1}{2}\mathcal U^{[+]}\,\Big\}\,
\psi(\xi)\,\rangle\\
%%%%
\delta\bar\Phi^{(q\bar q\rightarrow g\gamma)}(x{,}p_T)
=\pi\delta\bar\Phi_G^{(q\bar q\rightarrow g\gamma)}(x{,}p_T)
\propto\langle\,\overline \psi(0)\,
\Big\{\,\tfrac{1}{2}\frac{\tr[\mathcal U^{[\Box]}]}{N}\mathcal U^{[+]\dagger}
-\tfrac{1}{2}\mathcal U^{[+]\dagger}\,\Big\}\,
\psi(\xi)\,\rangle\\
%%%%
\delta\widehat\Gamma^{(qg\rightarrow q\gamma)}(z{,}k_T)
=\pi\delta\widehat\Gamma_{G_f}^{(qg\rightarrow q\gamma)}(z{,}k_T)
=\pi\delta\widehat\Gamma_{G_d}^{(qg\rightarrow q\gamma)}(z{,}k_T)
=0%\\
%\nonumber
\end{gather}
\end{small}

%\begin{center}
%$\boxed{\ qg{\rightarrow}q\gamma\ }$
%\end{center}
\begin{small}
\begin{gather}
\textstyle{\boxed{\ qg{\rightarrow}q\gamma\ }}\nonumber\\
%%%%
\delta\Phi^{(qg\rightarrow q\gamma)}(x{,}p_T)
=-\pi\delta\Phi_G^{(qg\rightarrow q\gamma)}(x{,}p_T)
\propto\langle\,\overline \psi(0)\,
\Big\{\,\tfrac{1}{2}\frac{\tr[\mathcal U^{[\Box]}]}{N}\mathcal U^{[-]}
-\tfrac{1}{2}\mathcal U^{[-]}\,\Big\}\,\psi(\xi)\,\rangle\\
%%%%
\delta\Delta^{(qg\rightarrow q\gamma)}(z{,}k_T)
=\pi\delta\Delta_G^{(qg\rightarrow q\gamma)}(z{,}k_T)
=0\displaybreak[0]\\
%%%%
\delta\Gamma^{(qg\rightarrow q\gamma)}(x{,}p_T)
\propto\langle\,\tr\Big[\,
F(0)\mathcal U^{[+]}F(\xi)
\Big\{\tfrac{1}{2}\mathcal U^{[-]\dagger}
{-}\tfrac{1}{2}\mathcal U^{[+]\dagger}\Big\}%\nonumber\\
%&\mspace{250mu}
+F(0)\mathcal U^{[-]}F(\xi)
\Big\{\tfrac{1}{2}\mathcal U^{[+]\dagger}
{-}\tfrac{1}{2}\mathcal U^{[-]\dagger}\Big\}\,\Big]\,\rangle\\
%%%%
\pi\delta\Gamma_{G_f}^{(qg\rightarrow q\gamma)}(x{,}p_T)
=\pi\delta\Gamma_{G_d}^{(qg\rightarrow q\gamma)}(x{,}p_T)
=0%\\
%\nonumber
\end{gather}
\end{small}

%\begin{center}
%$\boxed{\ qq{\rightarrow}qq\ }$
%\end{center}
\begin{small}
\begin{gather}
\textstyle{\boxed{\ qq{\rightarrow}qq\ }}\nonumber\\
%%%%
\delta\Phi^{(qq\rightarrow qq)}(x{,}p_T)
\propto\langle\,\overline \psi(0)\,
\Big\{\,\tfrac{3}{2}\frac{\tr[\mathcal U^{[\Box]}]}{N}\mathcal U^{[+]}
-\tfrac{1}{2}\mathcal U^{[\Box]}\mathcal U^{[+]}
-\tfrac{1}{2}\mathcal U^{[+]}-\tfrac{1}{2}\mathcal U^{[-]}\,\Big\}\,
\psi(\xi)\,\rangle\\
%%%%
\pi\delta\Phi_G^{(qq\rightarrow qq)}(x{,}p_T)
\propto\langle\,\overline \psi(0)\,
\Big\{\,\tfrac{1}{2}\mathcal U^{[\Box]}\mathcal U^{[+]}
-\tfrac{1}{2}\frac{\tr[\mathcal U^{[\Box]}]}{N}\mathcal U^{[+]}
-\tfrac{1}{2}\mathcal U^{[+]}+\tfrac{1}{2}\mathcal U^{[-]}\,\Big\}\,
\psi(\xi)\,\rangle%\\
%\nonumber
\end{gather}
\end{small}

%\begin{center}
%$\boxed{\ q\bar q{\rightarrow}q\bar q\ }$
%\end{center}
\begin{small}
\begin{gather}
\textstyle{\boxed{\ q\bar q{\rightarrow}q\bar q\ }}\nonumber\\
%%%%
\delta\Phi^{(q\bar q\rightarrow q\bar q)}(x{,}p_T)
=\pi\delta\Phi_G^{(q\bar q\rightarrow q\bar q)}(x{,}p_T)
\propto\langle\,\overline \psi(0)\,
\Big\{\,\tfrac{1}{2}\frac{\tr[\mathcal U^{[\Box]\dagger}]}{N}\mathcal U^{[+]}
-\tfrac{1}{2}\mathcal U^{[+]}\,\Big\}\,\psi(\xi)\,\rangle%\\
%\nonumber
\end{gather}
\end{small}

%\begin{center}
%$\boxed{\ qg{\rightarrow}qg\ }$
%\end{center}
\begin{small}
\begin{gather}
\textstyle{\boxed{\ qg{\rightarrow}qg\ }}\nonumber\\
%%%%
\delta\Phi^{(qg\rightarrow qg)}(x{,}p_T)
\propto\langle\,\overline \psi(0)\,
\Big\{\,\tfrac{1}{2}\frac{N^2{+}1}{N^2{-}1}\frac{\tr[\mathcal U^{[\Box]}]}{N}
\frac{\tr[\mathcal U^{[\Box]\dagger}]}{N}\mathcal U^{[+]}
-\tfrac{1}{2}\frac{N^2{+}1}{N^2{-}1}\mathcal U^{[+]}
+\tfrac{1}{2}\frac{\tr[\mathcal U^{[\Box]}]}{N}\mathcal U^{[-]}
-\tfrac{1}{2}\mathcal U^{[-]}\,\Big\}\,\psi(\xi)\,\rangle\\
%%%%
\pi\delta\Phi_G^{(qg\rightarrow qg)}(x{,}p_T)
\propto\langle\,\overline \psi(0)\,
\Big\{\,\tfrac{1}{2}\frac{\tr[\mathcal U^{[\Box]}]}{N}
\frac{\tr[\mathcal U^{[\Box]\dagger}]}{N}\mathcal U^{[+]}
-\tfrac{1}{2}\mathcal U^{[+]}
-\tfrac{1}{2}\frac{\tr[\mathcal U^{[\Box]}]}{N}\mathcal U^{[-]}
+\tfrac{1}{2}\mathcal U^{[-]}\,\Big\}\,\psi(\xi)\,\rangle\displaybreak[1]\\
%%%%
\delta\Gamma^{(qg\rightarrow qg)}(x{,}p_T)
\propto\langle\,\tr\Big[\,F(0)\mathcal U^{[+]}F(\xi)
\Big\{\tfrac{1}{2}\frac{\tr[\mathcal U^{[\Box]\dagger}]}{N}
\mathcal U^{[-]\dagger}
{-}\tfrac{1}{2}\mathcal U^{[-]\dagger}\Big\}
+F(0)\mathcal U^{[-]}F(\xi)
\Big\{\tfrac{1}{2}\frac{\tr[\mathcal U^{[\Box]}]}{N}\mathcal U^{[+]\dagger}
{-}\tfrac{1}{2}\mathcal U^{[+]\dagger}\Big\}\,\Big]\,\rangle\\
%%%%
\pi\delta\Gamma_{G_f}^{(qg\rightarrow qg)}(x{,}p_T)
\propto\langle\,\tr\Big[\,F(0)\mathcal U^{[+]}F(\xi)
\Big\{\tfrac{1}{2}\frac{\tr[\mathcal U^{[\Box]\dagger}]}{N}
\mathcal U^{[-]\dagger}
{-}\tfrac{1}{2}\mathcal U^{[-]\dagger}\Big\}
-F(0)\mathcal U^{[-]}F(\xi)
\Big\{\tfrac{1}{2}\frac{\tr[\mathcal U^{[\Box]}]}{N}\mathcal U^{[+]\dagger}
{-}\tfrac{1}{2}\mathcal U^{[+]\dagger}\Big\}\,\Big]\,\rangle\\
%%%%
\pi\delta\Gamma_{G_d}^{(qg\rightarrow qg)}(x{,}p_T)
\propto\langle\,\tr\Big[\,F(0)\mathcal U^{[+]}F(\xi)
\Big\{\tfrac{1}{2}\mathcal U^{[-]\dagger}
{-}\tfrac{1}{2}\frac{\tr[\mathcal U^{[\Box]}]}{N}
\mathcal U^{[+]\dagger}\Big\}
+F(0)\mathcal U^{[-]}F(\xi)
\Big\{\tfrac{1}{2}\mathcal U^{[+]\dagger}
{-}\tfrac{1}{2}\frac{\tr[\mathcal U^{[\Box]\dagger}]}{N}
\mathcal U^{[-]\dagger}\Big\}\,\Big]\,\rangle%\\
%\nonumber
\end{gather}
\end{small}

%\subsection*{$\bar qg\rightarrow\bar qg$}

%\begin{center}
%$\boxed{\ q\bar q{\rightarrow}gg\ }$
%\end{center}
\begin{small}
\begin{gather}
\textstyle{\boxed{\ q\bar q{\rightarrow}gg\ }}\nonumber\\
%%%%
\delta\Phi^{(q\bar q\rightarrow gg)}(x{,}p_T)
=\pi\delta\Phi_G^{(q\bar q\rightarrow gg)}(x{,}p_T)
\propto\langle\,\overline \psi(0)\,
\Big\{\,\tfrac{1}{2}\frac{\tr[\mathcal U^{[\Box]\dagger}]}{N}\mathcal U^{[+]}
-\tfrac{1}{2}\mathcal U^{[+]}\,\Big\}\,\psi(\xi)\,\rangle%\\
%\nonumber
\end{gather}
\end{small}

%\begin{center}
%$\boxed{\ gg{\rightarrow}q\bar q\ }$
%\end{center}
\begin{small}
\begin{gather}
\textstyle{\boxed{\ gg{\rightarrow}q\bar q\ }}\nonumber\\
%%%%
\begin{split}
\delta\Gamma^{(gg\rightarrow q\bar q)}(x{,}p_T)
&=-\pi\delta\Gamma_{G_f}^{(gg\rightarrow q\bar q)}(x{,}p_T)\\
%%%%
&\propto\langle\,\tr\Big[\,
F(0)\mathcal U^{[+]}F(\xi)
\Big\{\tfrac{1}{2}\frac{\tr[\mathcal U^{[\Box]\dagger}]}{N}
\mathcal U^{[-]\dagger}
{-}\tfrac{1}{2}\mathcal U^{[+]\dagger}\Big\}%\nonumber\\
%&\mspace{250mu}
+F(0)\mathcal U^{[-]}F(\xi)
\Big\{\tfrac{1}{2}\frac{\tr[\mathcal U^{[\Box]}]}{N}\mathcal U^{[+]\dagger}
{-}\tfrac{1}{2}\mathcal U^{[-]\dagger}\Big\}\,\Big]\,\rangle\end{split}\\
%%%%
\pi\delta\Gamma_{G_d}^{(gg\rightarrow q\bar q)}(x{,}p_T)
\propto\langle\,\tr\Big[\,
F(0)\mathcal U^{[+]}F(\xi)
\Big\{\tfrac{1}{2}\frac{\tr[\mathcal U^{[\Box]\dagger}]}{N}
\mathcal U^{[-]\dagger}
{-}\tfrac{1}{2}\mathcal U^{[-]\dagger}\Big\}
-F(0)\mathcal U^{[-]}F(\xi)
\Big\{\tfrac{1}{2}\frac{\tr[\mathcal U^{[\Box]}]}{N}\mathcal U^{[+]\dagger}
{-}\tfrac{1}{2}\mathcal U^{[+]\dagger}\Big\}\,\Big]\,\rangle%\\
%\nonumber
\end{gather}
\end{small}

%\begin{center}
%$\boxed{\ gg{\rightarrow}gg\ }$
%\end{center}
\begin{small}
\begin{gather}
\textstyle{\boxed{\ gg{\rightarrow}gg\ }}\nonumber\\
%%%%
\delta\Gamma^{(gg\rightarrow gg)}(x{,}p_T)
\propto\langle\,\tr\Big[\,
F(0)\mathcal U^{[+]}F(\xi)
\Big\{\tfrac{1}{2}\frac{\tr[\mathcal U^{[\Box]\dagger}]}{N}
\mathcal U^{[-]\dagger}
{-}\tfrac{1}{2}\mathcal U^{[+]\dagger}\Big\}
+F(0)\mathcal U^{[-]}F(\xi)
\Big\{\tfrac{1}{2}\frac{\tr[\mathcal U^{[\Box]}]}{N}\mathcal U^{[+]\dagger}
{-}\tfrac{1}{2}\mathcal U^{[-]\dagger}\Big\}\,\Big]\,\rangle\displaybreak[0]\\
%%%%
\pi\delta\Gamma_{G_f}^{(gg\rightarrow gg)}(x{,}p_T)
\propto
\langle\,\tr\Big[\,
\frac{1}{N^2}F(\xi)\mathcal U^{[\Box]\dagger}\mathcal U^{[+]\dagger}
F(0)\mathcal U^{[\Box]}\mathcal U^{[+]}
+F(0)\mathcal U^{[+]}F(\xi)\mathcal U^{[+]\dagger}
\frac{\tr[\mathcal U^{[\Box]}]}{N}\frac{\tr[\mathcal U^{[\Box]\dagger}]}{N}\nonumber\\
\mspace{457mu}
+F(0)\mathcal U^{[+]}F(\xi)
\Big\{\,\frac{N^2{+}4}{2N^2}\mathcal U^{[+]\dagger}
{-}\tfrac{1}{2}\frac{\tr[\mathcal U^{[\Box]\dagger}]}{N}
\mathcal U^{[-]\dagger}\,
\Big\}\nonumber\\
\mspace{457mu}
+F(0)\mathcal U^{[-]}F(\xi)
\Big\{\,\frac{N^2{+}2}{2N^2}\mathcal U^{[-]\dagger}
{-}\tfrac{1}{2}\frac{\tr[\mathcal U^{[\Box]}]}{N}
\mathcal U^{[+]\dagger}\,\Big\}\,\Big]\,\rangle\\
%%%%
\pi\delta\Gamma_{G_d}^{(gg\rightarrow gg)}(x{,}p_T)
=0
\end{gather}
\end{small}

\bibliographystyle{apsrev}
\bibliography{references}

\end{document}